\documentclass[twocolumn,preprintnumbers,amsmath,amssymb,aps,prb,longbibliography]{revtex4-1}
\usepackage{graphicx}
\usepackage{color}
 \usepackage{soul}

\begin{document}

\title{Perspective on Reversible to Irreversible Transitions in Periodic Driven Many Body Systems and Future Directions For Classical and Quantum Systems} 
\author{C. Reichhardt$^{1}$,
  Ido Regev$^{2}$,
  K. Dahmen$^{3}$, S. Okuma$^4$, and
  C. J. O. Reichhardt$^{1}$
}
\affiliation{
$^1$Theoretical Division,
Los Alamos National Laboratory, Los Alamos, New Mexico 87545 USA\\ 
$^2$Department of Solar Energy and Environmental Physics, Jacob Blaustein Institutes for Desert Research,
Ben-Gurion University of the Negev, Sede Boqer Campus 84990, Israel\\
$^3$Department of Physics, University of Illinois at Urbana-Champaign,
1110 West Green Street, Urbana, IL 61801, USA\\
$^4$Department of Physics, Tokyo Institute of Technology, 2-12-1,
Ohokayama, Meguro-ku, Tokyo 152-8551, Japan
} 

\date{\today}
\begin{abstract}

Reversible to irreversible (R-IR) transitions
arise in numerous 
periodically driven collectively interacting systems
that, after a certain number of driving cycles,
organize
into a reversible state where the particle trajectories
repeat during every or every few cycles.
On the irreversible side of the transition,
the motion is chaotic.
R-IR transitions were first
systematically studied for periodically sheared
dilute colloids, and have now 
been found in a wide variety of
both soft and hard matter periodically
driven systems, including amorphous solids, crystals,
vortices in type-II superconductors,
and magnetic textures.
It has been shown that in several of these systems, the
transition to
a reversible state is
an absorbing phase transition 
with
a critical divergence in the 
organization time scale at the transition.
The same systems are capable of
storing multiple
memories and may
exhibit return point memory.
We give an overview of R-IR transitions
including recent advances in the field, and
discuss how the general framework of R-IR transitions
could be
applied 
to a much broader class 
of nonequilibrium systems
in which periodic driving occurs, including not only
soft and hard condensed matter systems, but also
astrophysics, biological systems, and social systems.
In particular, some likely candidate systems are
commensurate-incommensurate states,
systems exhibiting hysteresis or 
avalanches,
nonequilibrium pattern forming states,
and other systems with absorbing phase transitions.
Periodic driving could be applied to hard condensed matter
systems to see if organization into reversible states
occurs for
metal-insulator transitions,
semiconductors, electron glasses, electron nematics, cold atom systems,
or Bose-Einstein condensates.
R-IR transitions could also be examined
in dynamical systems where synchronization or phase locking occurs.
We also discuss
the possibility of using complex periodic driving,
such as changing drive directions or using multiple
frequencies,
to determine whether these systems
can still organize to reversible states or retain complex multiple memories.
Finally, we describe features of
classical and quantum time crystals that could suggest the
occurrence of R-IR transitions in these systems.
\end{abstract}
\maketitle

\section{Introduction}

Driven many-body deterministic nonlinear systems generally 
exhibit disordered or chaotic dynamics, as found
in turbulence \cite{Deissler86},
particle flow over disordered media \cite{Fisher98},
plasmas \cite{Escande16},
sheared materials \cite{Mason96,Sierou04},
granular matter \cite{Hill99},
earthquakes \cite{Huang90},
gravitational systems \cite{Lecar01},
and biological systems \cite{Toker20}.
Chaotic dynamics can arise even in systems with only
three
degrees of
freedom \cite{Lorenz63,Ott93} so it can be expected that driven 
many body disordered systems with hundreds or thousands of degrees of 
freedom will generally exhibit fluctuating or chaotic dynamics. 
Recently, a growing number of many body systems
have been shown to exhibit a transition 
from
time periodic or reversible motion
to chaotic irreversible motion
under oscillatory
driving
\cite{Pine05,Corte08,Mangan08,Fiocco13,Regev13,Royer15,Keim14,Jana17,Keim19,Nagasawa19,Wilken20,Khirallah21,Maegochi21,Galloway22,Denisov2015}.
In these studies, the particle positions are compared from
one driving cycle to the next.
In the chaotic phase, the particles do not return to
the same positions and undergo diffusive motion away from the
initial positions over many driving cycles.
For certain driving amplitudes
or system parameters,
however,
the particles can organize over many cycles into
a reversible state in which they return to the same
positions after every or every few cycles,
and the long time diffusive behavior is lost.

Reversible behavior in viscosity-dominated flows was
famously demonstrated
by G.~I.~Taylor \cite{Taylor67} using a two cylinder setup in which the inner
cylinder is rotated multiple times and then rotated back.
Pine {\it et al.} used the same shearing geometry and viscous
fluid as Taylor but considered the case where there are additional
colloidal particles
in the fluid that can collide with each other,
so that any irreversible behavior would be due to the
particle collisions rather than the fluid itself \cite{Pine05}.  
The simple periodically sheared dilute colloidal particles experiment
allows for the  
systematic study
of transitions from irreversible to reversible motion
in many-body systems.
Pine {\it et al.} showed
that there is a critical strain amplitude below 
which the
steady state behavior
is reversible and above which the behavior
becomes diffusive or irreversible.
The colloids are in a viscous fluid and are large enough that
thermal effects are negligible;
since they are also electrically neutral,
the only interactions capable of producing an irreversible state
are contact forces between the 
colloids during collisions.
During the initial shear cycles,
particles move by different amounts and
some particles collide with one another.
In
a
steady irreversible state,
collisions
occur
during each cycle
and cause
the particles to wander away from their initial positions in
a Brownian-like diffusion, where the
distances
traveled along the shear direction, $\langle x^2\rangle$,
and perpendicular to the shear direction,
$\langle z^2\rangle$, increase linearly with time.

In further work,
Corte {\it et al.}~\cite{Corte08}
studied the number of
drive cycles required
for the system to reach a reversible state 
from a randomized initial state.
In particular, during the first cycles the motion
is chaotic, but
after many cycles the system
may organize to a state where collisions are absent.
The
number of transient cycles $n_\tau$ spent reaching the reversible
state
diverges
at a critical density or  
at a critical shear amplitude
$\gamma_c$ according to
a power law $n_\tau = |\gamma -\gamma_{c}|^\nu$
\cite{Corte08}.
This critical behavior suggests that the
transition to
the reversible state is a nonequilibrium 
phase transition,
and the observed exponents $\beta = 0.45(2)$ and $\nu_{\parallel} = 1.33(2)$ are
similar to the critical exponents
of the two-dimensional
directed percolation (DP)
universality class,
namely $\beta = 0.584(4)$ and $\nu_{\parallel} = 1.295(6)$
\cite{Hinrichsen00}.  
Although extensive theoretical studies of DP 
transitions have been performed,
clear observations of these transitions in experimental
systems have only been obtained relatively recently
\cite{Takeuchi09,Lemoult16,Marcuzzi16}.

Since R-IR
transitions were first observed
in dilute systems,
it might seem reasonable to imagine that these transitions
only arise under specialized circumstances
where the interactions are
of sufficiently short range or 
the system
is sufficiently dilute that
it is possible to reach
a state where collisions never occur.
This would imply 
that it would be difficult to find reversible states 
in disordered strongly coupled systems
where
the particles are always in contact or
where long range interactions are relevant; 
however, R-IR
transitions
have in fact been observed
for periodically driven granular systems \cite{Royer15}
where the particles are always in contact
as well as
for vortices in type-II superconductors \cite{Mangan08,Maegochi21}
at densities for which
the vortices are strongly interacting. 
One of the most extensively  studied strongly
interacting systems exhibiting
an IR-R transition
is periodically sheared amorphous solids
\cite{Fiocco13,Regev13,Keim14,Jana17,Nagasawa19,Khirallah21,Galloway22,Regev15,Fiocco14,Keim21}.
Here the particles or atoms
are always in contact and long range strain fields are present.
Reversible behavior is expected to occur in solids
when the strain amplitude is small
so that the system behaves
completely elastically;
however, the R-IR transitions
in the amorphous solids 
appear well into regimes where
strong plastic deformations occur in which
particles are exchanging neighbors.
In general,
in strongly coupled systems
the particle trajectories during
reversible states
show more complex loops
or return to the same position not after every drive cycle but
after multiple drive cycles, such as in examples where
the motion repeats every
six cycles \cite{Regev13,Nagasawa19,Keim21,Lavrentovich17}. 
Plasticity is thought to be an inherently irreversible process; however,
these works indicate that it is possible for
plastically deforming reversible regimes to emerge,
suggesting that similar transitions to reversible
states could arise
in many other strongly coupled systems.
There have now been several other
studies of many body
interacting systems undergoing IR-R transitions
including crystals \cite{Ni19},
systems with quenched disorder \cite{Mangan08,Maegochi21,Okuma11},
chiral active matter \cite{Lei19,Reichhardt19},
skyrmions \cite{Brown19}, and magnetic materials \cite{Basak20}.
Additionally, it has been shown that in the reversible phase,
the system can be trained to store multiple memories \cite{Keim19,Fiocco14}.

In this paper we give an overview of 
work on reversible-irreversible transitions in both
dilute and strongly coupled systems,
and discuss how ideas
about R-IR transitions
could be more broadly applied to other areas
including polymeric soft matter, pattern forming systems,
and frictional systems. 
The R-IR transition could be induced via
global driving such as a shear or external field,
or via local driving such as
the periodic motion of a local probe. 
We also discuss classes of active matter systems that could also
exhibit R-IR transitions in the limit
where thermal effects are negligible
but the activity can be treated as periodic.
Most existing studies
involved a periodic drive applied
in a single direction at one frequency;
however, much more complicated periodic drives could be applied
with multiple frequencies or changing directions, opening a new
area of investigation.
Such systems
might organize to more complex reversible states and 
could retain a memory of past drive protocols as the driving is
changed from more to less complex.
We also discuss possible applications of R-IR
transitions to a wider 
number of hard condensed matter systems under periodic driving such as
metal-insulator systems, charge ordering
systems, Bose-Einstein condensates, superfluids,
sliding charge density waves, Wigner crystals,
systems showing nonlinear transitions,
classical time crystals,
and numerous magnetic textures including magnetic domain walls
and magnetic bubbles.
These types of systems can be subjected to numerous forms of
periodic driving, 
such as electric currents, magnetic fields, or optical excitations.
The most promising quantum systems for observing R-IR transitions are the
time crystals \cite{Sacha18},
which already show a number of features
consistent with a transition from a chaotic state to
a time periodic state.
There is also a
broader class of nonlinear coupled
systems such as coupled oscillators
or networks that have been shown to exhibit synchronization
and phase locking effects \cite{Strogatz00,Pikovsky01} and
nonequilibrium pattern formation \cite{Cross93}.
Synchronization effects can occur in many body 
coupled systems when the many degrees of freedom become coupled
so that the response looks like that of a few body system, and
we argue that such transitions could be viewed as R-IR
transitions, suggesting that  
these systems
could also be candidates for exhibiting
transitions into absorbing states.

\section{Sheared Systems}

\begin{figure}
\includegraphics[width=3.5in]{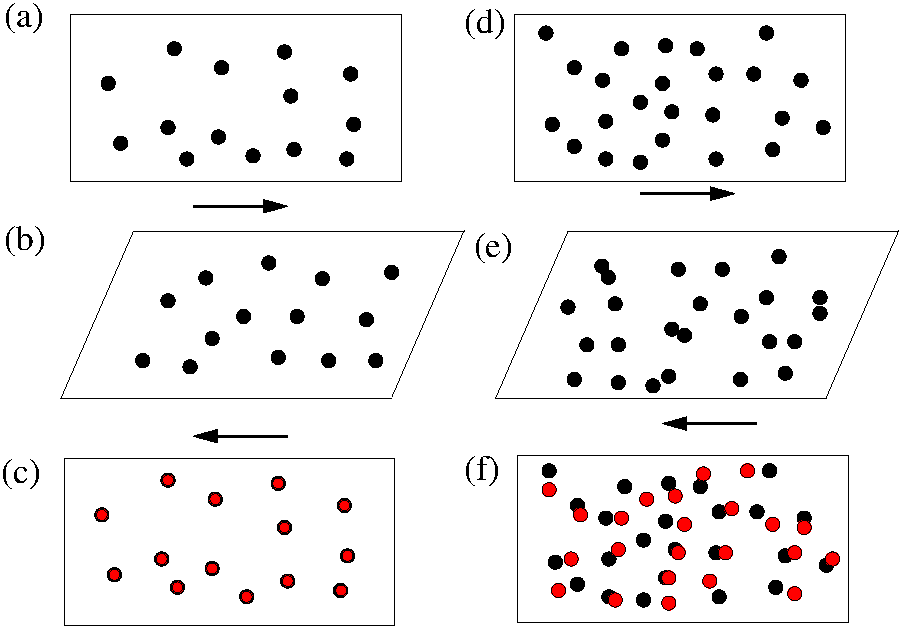}
\caption{
A schematic of the system
studied by Pine {\it et al.} \cite{Pine05}, consisting
of a dilute suspension of colloids (black disks) 
subjected to periodic shearing. Arrows indicate the shearing direction.
In (a) and (b), the system is in the dilute limit, and the comparison
of the starting (black) and ending (red) positions of the particles
shown in (c) indicates reversible behavior.
In (d) and (e), the system is in the dense limit,
the particles collide repeatedly under shear, and the
image of starting and ending positions in (f) indicates
the occurrence of irreversible motion.
}
\label{fig:1}
\end{figure}

\begin{figure}
\includegraphics[width=3.5in]{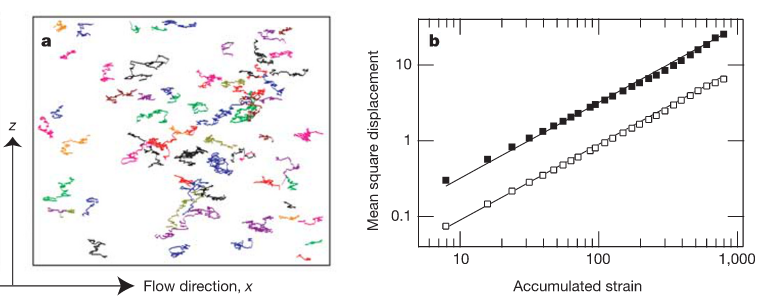}
\caption{(a) Particle trajectories measured experimentally for the
sheared colloid system in Ref.~\cite{Pine05} in the irreversible or
chaotic regime.  
(b) The corresponding mean square particle displacements in the direction
parallel, $\langle x^2\rangle$ (filled squares), and
perpendicular, $\langle z^2\rangle$ (open squares), to the
drive, showing diffusive motion.
Reprinted by permission from: Springer Nature,
D.~J.~Pine {\it et al.}, ``Chaos and threshold for irreversibility in sheared
suspensions'', Nature (London) {\bf 438}, 997 (2005).
}
\label{fig:2}
\end{figure}

The dilute colloid experiments mentioned above that were
performed by Pine {\it et al.} \cite{Pine05} were
the first studies of transitions 
from irreversible to reversible states.
In Fig.~\ref{fig:1},
representative configurations show the situation in the reversible
versus irreversible steady states
for systems with
low [Fig.~\ref{fig:1}(a)] and high [Fig.~\ref{fig:1}(b)]
colloid
densities
subject to the same amplitude of shearing.
Comparison of the
starting
(black dots)
and ending
(red dots)
positions of the particles, shown
in Fig.~\ref{fig:1}(c,f) for the two densities, reveals that
the motion is reversible in the low density system
where collisions do not occur and irreversible
in the higher density system where collisions are ongoing.
Figure~\ref{fig:2}(a) shows
the experimentally measured particle positions from
Ref.~\cite{Pine05} 
at the end of
each cycle over multiple cycles in the irreversible regime.
Here the
dynamics
is Brownian-like,
and
the particles gradually diffuse away from their initial  positions.
In Fig.~\ref{fig:2}(b), the corresponding measured mean square displacements
$\langle x^2\rangle$ in the direction of drive
versus accumulated stain
increases linearly
as expected for Brownian motion.
The displacements $\langle z^2\rangle$ in the direction perpendicular
to the drive
show the same behavior but have a lower value.

\begin{figure}
\includegraphics[width=3.5in]{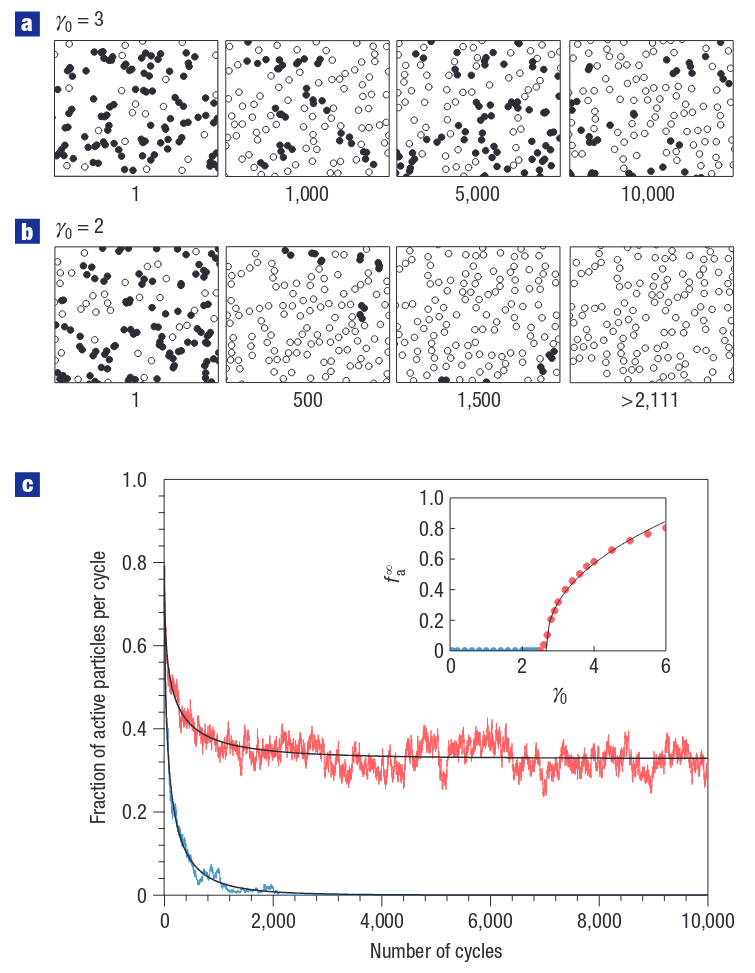}
\caption{
(a,b) Snapshots from simulations in Ref.~\cite{Corte08} of a hard disk
model of the experiments in Fig.~\ref{fig:2} as a function of time
for different strain amplitudes $\gamma_0$. Black disks are moving
irreversibly between cycles and open disks are returning to the same
position between cycles.
(a) $\gamma_0=3$
in the irreversible regime. (b) $\gamma_0=2$ in the reversible regime.
(c) For the same system, the fraction of particles that are active (moving
irreversibly) in each cycle as a function of the number of shear cycles
applied at $\gamma_0=3$ (red) and $\gamma_0=2$ (blue). The inset shows the
steady state fraction of active particles as a function of $\gamma_0$.
Reprinted by permission from: Springer Nature,
L.~Cort{\' e} {\it et al.}, ``Random organization in periodically driven
systems'', Nature Phys. {\bf 4}, 420 (2008).
}
\label{fig:3}
\end{figure}

In general, for
any random initial condition in sheared colloidal systems,
during the first few cycles the particles
do not return to their original positions
after each cycle. Instead,
over time the particles either organize to a
reversible state or remain in an irreversible state.
Corte {\it et al.} \cite{Corte08} used a combination of simulation and
experiments to further explore
the system studied by
Pine {\it et al.} \cite{Pine05},
and applied different
strain amplitudes to
identify the manner in which
the system reaches a steady irreversible or reversible state.
Figure~\ref{fig:3}(a,b)
shows the positions of the active particles, or particles that did
not return to their original positions after each cycle,
for a simulation of two-dimensional (2D) sheared hard disks
at two different strain amplitudes
of $\gamma_{0} = 3.0$ [Fig.~\ref{fig:3}(a)]
and $\gamma_0=2.0$ [Fig.~\ref{fig:3}(b)] \cite{Corte08}.
At the end of the first shear cycle,
there are numerous active particles in each case, but over time
the system reaches a steady state
that is either irreversible with a finite number of active particles,
as shown for $\gamma_0 = 3.0$
in Fig.~\ref{fig:3}(a),
or reversible with no active particles,
as shown for $\gamma_0 = 2.0$ in Fig.~\ref{fig:3}(b).
A time series of the fraction of active particles as a function of
shear cycle number appears in 
Fig.~\ref{fig:3}(c) for the same two strain amplitudes.
After 2500 cycles, there are no active particles
remaining in the $\gamma_0 = 2.0$ sample,
but the activity in the
$\gamma_{0} = 3.0$ sample
plateaus at a finite
steady state value where
about 39\% of the particles remain active.
The inset in Fig.~\ref{fig:3}(c) shows the
steady state active particle fraction versus strain amplitude,
indicating that there is a transition from 
irreversible (active)  to reversible (nonactive) 
behavior near a critical strain of $\gamma^c_0 = 2.66$.

\begin{figure}
\includegraphics[width=3.5in]{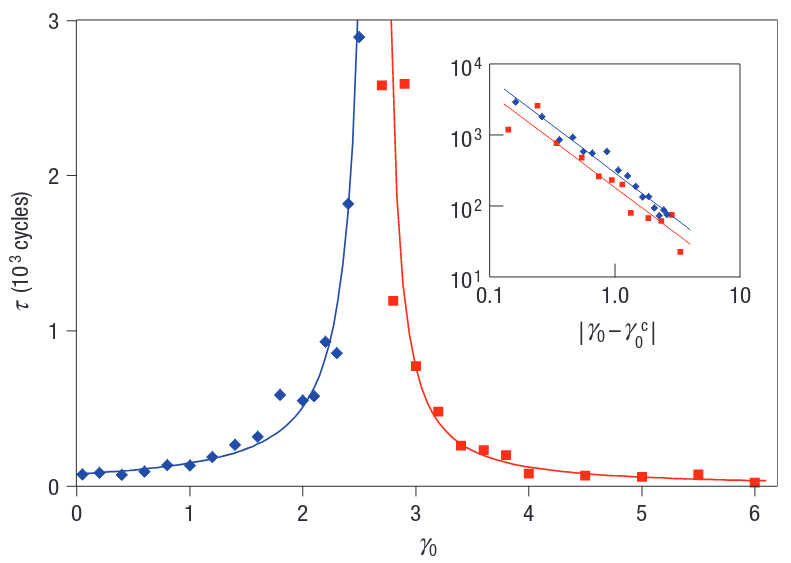}
\caption{
For the simulation of sheared 2D disks
from Fig.~\ref{fig:3}, a plot of the time $\tau$ to reach a steady
state versus shear amplitude $\gamma_0$ for the reversible (blue diamonds)
and irreversible (red squares) regimes. The inset shows the same data as
a power law plot of $\tau$ versus
$|\gamma_0-\gamma_0^c|$, where the critical shear amplitude
is $\gamma_0^c=2.66 \pm 0.05$. The lines indicate fits of the data
to $\tau \propto |\gamma_0-\gamma_0^c|^\nu$ with
$\nu =1.33\pm0.02$.
Reprinted by permission from: Springer Nature,
L.~Cort{\' e} {\it et al.}, ``Random organization in periodically driven
systems'', Nature Phys. {\bf 4}, 420 (2008).
}
\label{fig:4}
\end{figure}

By fitting the curves in Fig.~\ref{fig:3}(c) to a stretched exponential,
Corte {\it et al.} \cite{Corte08}
obtained the mean time $\tau$ required to reach a steady
reversible or irreversible state, plotted in
Fig.~\ref{fig:4} as a function of strain amplitude $\gamma_0$.
At the critical amplitude $\gamma^c_0$, $\tau$ diverges as
a power law according to
\begin{equation} 
\tau \propto |\gamma_{0} -\gamma^{c}_{0}|^{-\nu} \ .
\end{equation}
In the 2D simulations, it was found that $\nu \approx 1.33$ on both sides of the
transition, while the three-dimensional (3D) experiments produced a similar
divergence with $\nu = 1.1$.
This transition has the hallmarks
of what is known as an absorbing phase transition,
which appears in 
nonequilibrium systems and often
falls in the directed percolation (DP) universality class
\cite{Hinrichsen00,Menon09}.
The irreversible state can be viewed as
a dynamically fluctuating state in which
the particles continue to exchange positions
and long time diffusion is occurring.
In contrast,
the reversible state corresponds to
the absorbed state where the fluctuations are lost
and the behavior becomes
completely time repeatable, indicating that the system
is
dynamically frozen or trapped in a limit cycle.

Corte {\it et al.} \cite{Corte08}
called the transition into the reversible state
``random organization'' since the
particles form a random configuration in which collisions do
not occur.
Since that time there have been further studies of
random organization
in these sheared dilute colloidal systems \cite{Jeanneret14},
as well as a number of studies indicating that
randomly organized states
near the
critical threshold are hyperuniform
\cite{Hexner15,Weijs15,Tjhung15,Hexner17,Lei19a,Lei19,Wilken20,Zheng21}.
In hyperuniform states,
there are no large fluctuations at
long length scales.
Periodic crystals are hyperuniform by definition; however,
certain random structures also have
hyperuniform properties \cite{Torquato03},
and there is ongoing work to understand 
the conditions under which random hyperuniform states can occur. 
In a random organized state,
the particles form configurations where collisions are
absent. This increases the
average distance between particles and reduces 
large
local variations in the particle positions,
thereby diminishing
large density fluctuations
and giving a more 
uniform density that extends out to long length scales. 
An open question is whether all systems near a R-IR transition
exhibit hyperuniformity, or whether this is 
a property found only in systems that are dilute
or that have short range interactions.

\begin{figure}
\includegraphics[width=3.5in]{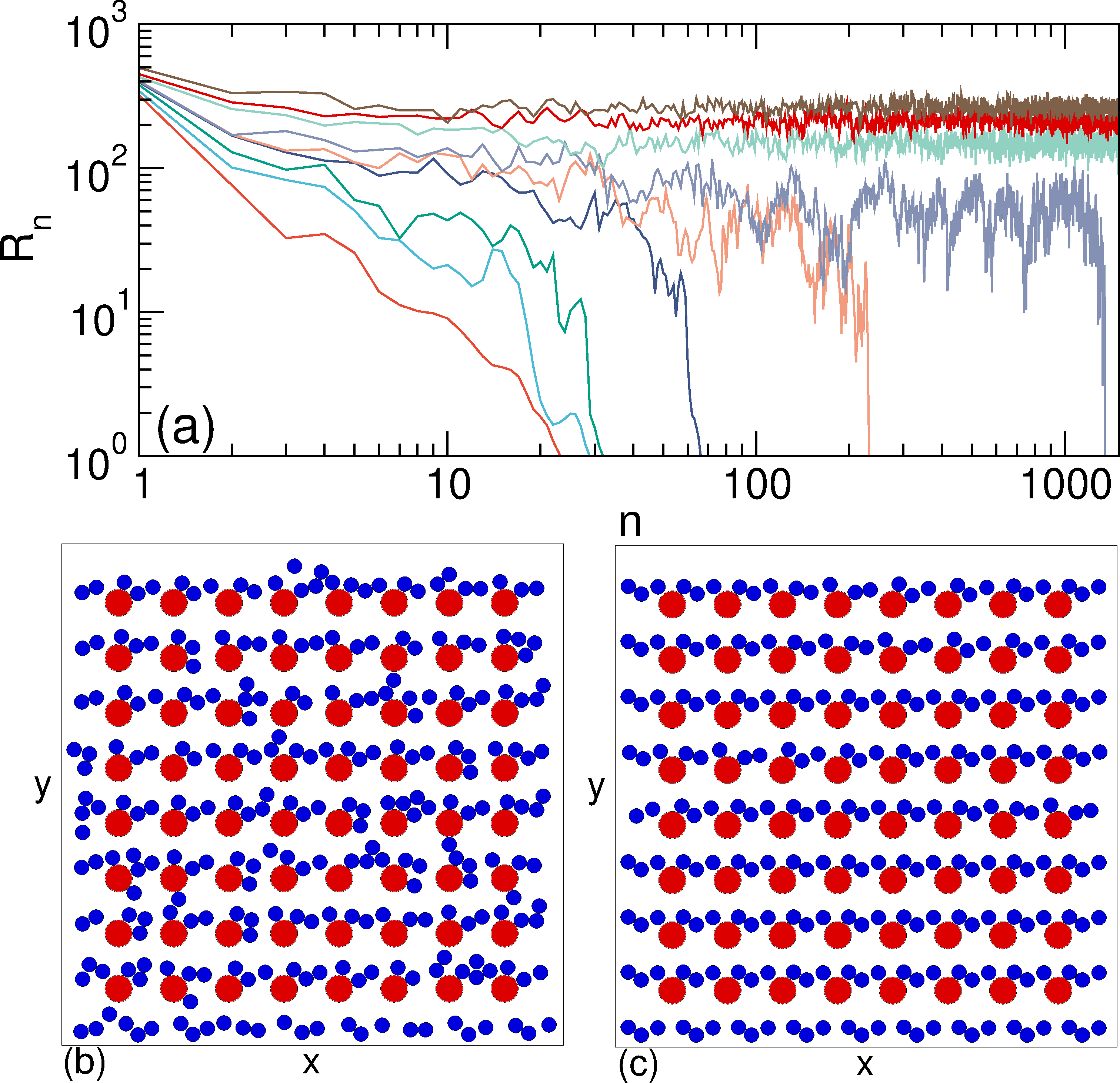}
\caption{
Simulations of cyclically driven disks in a periodic array of
obstacles from Ref.~\cite{Reichhardt22}. (a) The number $R_n$ of active
or irreversibly moving disks versus cycle number $n$ for different
disk densities ranging from $\phi=0.335$ (bottom) to $\phi=0.3962$ (top)
under shearing with an amplitude of $A=0.031623$ at an angle of
$\theta = 18.435^\circ$ from the x-axis symmetry direction of the
obstacle array.
(b,c) Images of the disk locations (blue) and obstacles (red) in
a portion of the sample for the same driving amplitude and direction
as in panel (a). (b) An irreversible state at $\phi=0.3962$.
(c) A reversible state at $\phi=0.3716$.
Reprinted from C. Reichhardt and C.J.O. Reichhardt, J.~Chem.~Phys.~{\bf 156},
124901 (2022) with the permission of AIP Publishing.
}
\label{fig:5}
\end{figure}

There has also been some work on ordered pattern formation in dilute
systems, where the R-IR transition overlaps with a
disorder (irreversible) to order (reversible) transition.
One example is hard disk colloidal particles
interacting with a periodic array of obstacles
\cite{Reichhardt22}.
Figure~\ref{fig:5}(a) shows the fraction $R_n$ of
active particles at the end of each cycle for
periodically driven disks moving through
an ordered array of posts at a fixed
drive amplitude for different disk densities $\phi$.
When $\phi > 0.3716$, the system remains in
an irreversible state, while
for $\phi \leq 0.3716$, the disks organize into a reversible state.
Figure~\ref{fig:5}(b,c) illustrates
the particle positions
in the steady state, which is
irreversible and disordered in
Fig.~\ref{fig:5}(b)
at $\phi = 0.3962$,
and reversible and ordered at
$\phi=0.3716$ in Fig.~\ref{fig:5}(c).
In Ref.~\cite{Reichhardt22}, the authors
also found a power
law divergence of the time required to reach
the reversible state.
The power law exponents are similar to those observed in 
2D random organization systems,
suggesting that the periodically driven disks fall
in the same universality class as those systems.
Another interesting feature of
the reversible state in Fig.~\ref{fig:5}(c) is that
collisions are not absent but instead
the particles collide with the obstacles in a repeating pattern.
Pattern formation in a reversible state was also studied for bidisperse
systems in which half of the particles move in circles and the other
half do not, where it was shown that there is
a transition from a mixed fluid to
a pattern forming phase separated reversible state \cite{Reichhardt19}.

\begin{figure}
\includegraphics[width=3.5in]{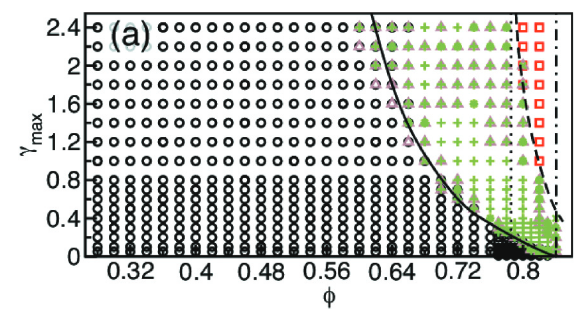}
\caption{
R-IR transition in simulations of a 2D bidisperse assembly of hard
disks from Ref.~\cite{Schreck13}. The dynamic phases are plotted as a
function of the strain amplitude $\gamma_{\rm max}$ versus the
disk density $\phi$. Black symbols indicate reversible states where
the disks return to their original locations after a single cycle.
For the green symbols, the states are loop reversible with complicated
disk trajectories that repeat after one or more periods. In the region
with red symbols, the steady state is irreversible.
Reprinted with permission from C.~F.~Schreck {\it et al.}, Phys. Rev. E
{\bf 88}, 052205 (2013).
Copyright by the American Physical Society.
}
\label{fig:6}
\end{figure}

\section{Condensed systems}
We next consider R-IR transitions
in systems close
to or just at jamming, as well as for systems deep in the
jamming phase and amorphous solids. 
In dilute systems, the irreversible state generally shows
a liquid structure and the particles 
do not form a solid.
At the transition into the absorbing state,
the particles either experience a small number of
repeating collisions or have no collisions at all.
As a result, it might be
assumed that R-IR transitions
are limited only to systems
that are dilute or have contact interactions,
providing the particles with enough space to rearrange and
organize into a reversible state. Below we show that this is not
the case.

\subsection{Reversibility near Jamming and Packing}
Schreck {\it et al.} \cite{Schreck13}
studied a granular matter version of R-IR transitions for
a periodically driven 2D bidisperse disk assembly,
and focused on the formation of a reversible state below the jamming
transition, $\phi<\phi_J=0.84$.
Well below jamming, the system organizes into a
reversible state where there are no collisions.
Schreck {\it et al.} term this a ``point reversible'' state, and it is
the same as a random organization state.
For higher densities, the system can still
organize to a reversible state,
but the disk trajectories become much more complex
and involve some collisions.
Schreck {\it et al.} named these
``loop reversible'' states since
the reversible orbits form loop structures instead of
straight lines.
At densities near jamming, an irreversible steady state emerges.
Figure~\ref{fig:6} shows the phase diagram as a function of
strain amplitude $\gamma_{\rm max}$ and disk density $\phi$ from
Ref.~\cite{Schreck13}, where the black region is point reversible, the
green region is loop reversible, and the irreversible states are
colored in red. Schreck {\it et al.} also obtained
similar results in 3D systems.
Recent studies by Gosh {\it et al.} 
on disk packings
showed that the transition
to a reversible
state
coincides with a transition to a crystalline state,
which is interesting
because it was not expected {\it a priori} that these transitions would
occur at the same point \cite{Ghosh22}.

Both experiments and simulations have been performed for R-IR transitions
in granular matter as a function of
varied shear amplitude \cite{Slotterback12} and friction \cite{Royer15}.
Other works
address transitions to reversible states or
random organization at the approach to
jamming or random close packing \cite{Wilken21},
as well as possible ways to
connect jamming and yielding in a unified framework \cite{Milz13,Das20,Ness20}.
Open questions for systems transitioning between jammed and unjammed
states include
how the nature of the trajectories changes across jamming
and whether there could be
different types of absorbing transitions.
Other effects to consider would be
adding quenched disorder sites
to a jamming system \cite{Reichhardt12,Graves16} in order to determine how
the R-IR transition is affected, or to study whether R-IR
transitions change in the
presence of Griffiths \cite{Griffiths69} or Gardner transitions \cite{Charbonneau15}.

\subsection{Amorphous Solids}

\begin{figure}
\includegraphics[width=0.5\textwidth]{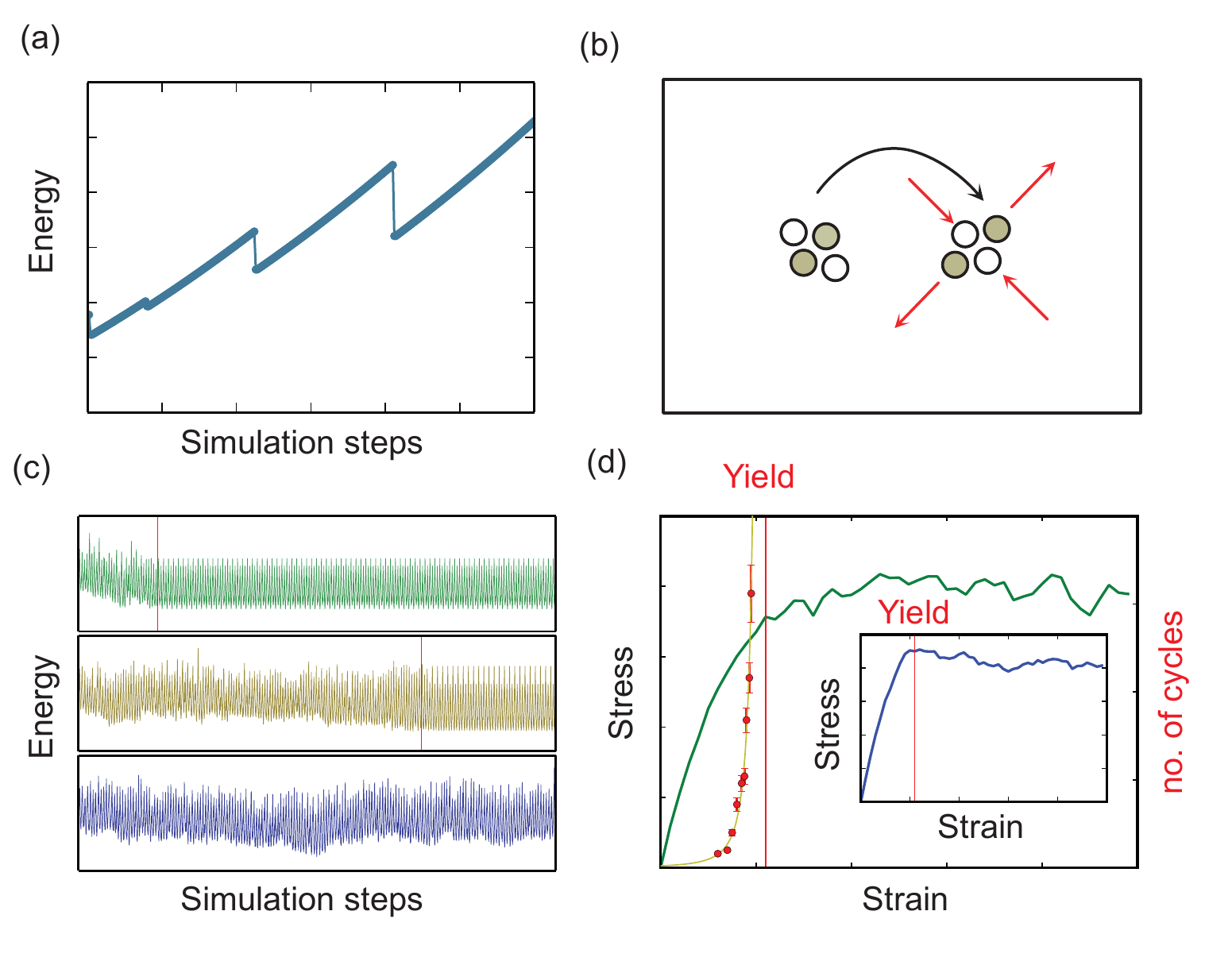}%
\caption{
(a)
The energy as a function of simulation steps for an athermal quasi-static
simulation of an amorphous solid. The discontinuous drops in the energy occur
due to plastic rearrangements.
Reprinted under CC license from I. Regev {\it et al.}, Nature Commun. {\bf 6}, 8805 (2015).
(b)
An illustration of particle motion in the most fundamental plastic rearrangement event, the soft-spot.
(c) 
Three different potential energy time series for three different maximal strain amplitudes, which increase from top to bottom. The red lines mark the onset of repetitive behavior or the formation of a limit cycle.
(d)
Stress-strain curve under linear shear (green line). The red vertical line marks the R-IR transition, while the red data points indicate the number of cycles required to reach a limit cycle under oscillatory shear. The inset shows a similar stress-strain curve obtained for different initial conditions.
Reprinted with permission from I.~Regev {\it et al.}, Phys. Rev. E
{\bf 88}, 062401 (2013).
Copyright by the American Physical Society.
}
\label{fig:7}
\end{figure}

For
systems such as solids or glasses well above the jamming density
subjected to shear,
it is known that at small strains,
the response is elastic and plastic rearrangements do not
occur,
while for intermediate strains,
plastic events start to appear, and for
even higher strains the system exhibits plastic yielding.
Currently there is a considerable amount of
work on elucidating the nature of the yielding
transition in sheared solids and glasses, 
understanding the shape of the stress-strain curve,
and determining the way in which
shear response and the yielding transition
depend on how the system is prepared.
Figure~\ref{fig:7}(a) shows the potential energy of an amorphous solid subject to an athermal strain increase every simulation step. 
The energy increases parabolically, as is expected from a rigid elastic material, and decreases discontinuously when a plastic rearrangement of the particles occurs. The most basic plastic rearrangement involves a change of nearest neighbors called a ``soft-spot''\cite{Argon79,Maloney06,Falk98,Manning11} that is illustrated in Fig.~\ref{fig:7}(b). Most plastic events involve several soft-spots, as will be discussed below.
Amorphous and crystalline systems
can also be subjected to
periodic shearing,
and for strains where the response is 
plastic under unidirectional shear, it could be assumed
that the system would only exhibit irreversible states; 
however, Fiocco {\it et al.} \cite{Fiocco13}, Regev {\it et al.} \cite{Regev13} and Priezjev \cite{Priezjev13} studied 3D and 2D model glasses subject to cyclic shear and showed the existence of a transition from reversible to irreversible dynamics at a critical strain amplitude. Priezjev \cite{Priezjev13} demonstrated that, at a finite temperature, there is a transition from an almost periodic, subdiffusive regime, to a diffusive regime. 
Fiocco {\it et al.} \cite{Fiocco13} and Regev {\it et al.} \cite{Regev13} used athermal quasi-static simulations to show that below the critical point the dynamics is exactly periodic and particles repeat the same positions after each cycle.
Regev {\it et al.} \cite{Regev13} found that the number of cycles needed to reach a limit-cycle diverges at this point.
In their study, Fiocco {\it et al.} \cite{Fiocco13} also showed that the post-yield dynamics involves a loss of memory of the initial configuration.

\begin{figure}
\includegraphics[width=3.5in]{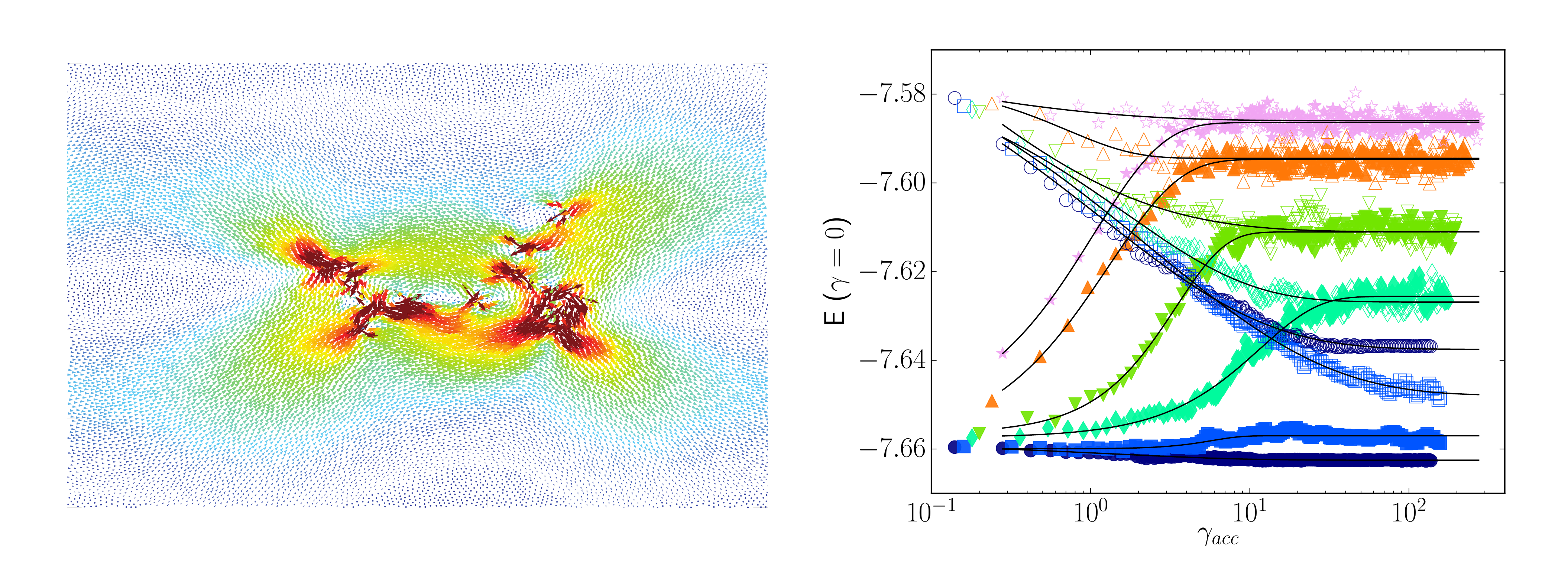}
\caption{
(a) 
A repeating avalanche in the reversible state of the system from
Fig.~\ref{fig:7}. Arrows and colors
indicate the direction and magnitude, respectively, of the
displacements during the avalanche motion.
Reprinted under CC license from I. Regev {\it et al.}, Nature Commun. {\bf 6}, 8805 (2015).
(b)
The potential energy $E$ per particle in the steady state under
zero strain ($\gamma=0$) for different values of maximum shear
strain $\gamma_{\rm max}$ increasing from bottom to top under 
temperatures $T=1.0$ (open symbols) and $T=0.466$ (closed symbols).
Reprinted with permission from D.~Fiocco {\it et al.}, Phys. Rev. E
{\bf 88}, 020301 (2013).
Copyright by the American Physical Society.
}
\label{fig:8}
\end{figure}

Figure~\ref{fig:7}(c)
shows the potential energy as a function of strain cycles for
increasing strain amplitudes. At the lowest amplitude, on the top portion
of the panel,
the system is initially in 
an irreversible state and settles after a short transient
into a reversible state,
as indicated by the transitions
from a fluctuating non-repeating signal to a periodic signal.
As the strain amplitude increases, it takes longer for the
system to reach a reversible state. Remarkably, even
in the reversible states the
system shows large scale {\it reversible} plastic deformations.
Figure~\ref{fig:8} shows an example of a
reversible plastic avalanche event for
the system
from Fig.~\ref{fig:7} \cite{Regev15}.
Figure~\ref{fig:7}(d) illustrates the 
number of cycles required to reach a reversible state
as a function of strain amplitude along with the stress-strain curve.
At yielding, marked by the red vertical line,
there is a divergence in the time needed to reach the
reversible state.
This work also showed that there is
a power-law divergence in the time scale to reach the reversible
state as a function of strain amplitude; however,
the observed critical exponent
$\nu \approx 2.6$ differs from
the value $\nu \approx 1.33$ obtained by Corte {\it et al.} \cite{Corte08},
suggesting that the transition is in a different universality class.
There are several possible reasons why this might be the case.

For a reversible state in dilute systems, there
are no collective effects since the particle-particle contacts vanish,
whereas in the amorphous system, the reversible
plastic events indicate that there is strong coupling among the
particles,
implying that there may be a dynamical length
scale present and that 
the microscopics of the reversible state are different in the
amorphous and dilute systems.
Another feature in the work of Regev
{\it et al.} \cite{Regev13} is that
the divergence in the reorganization
time scale correlates well with the point at which a yielding
transition appears,
so that
below yielding,
the oscillating drive creates a reversible
plastic state,
while above yielding,
the system can never reach a reversible state.

\begin{figure*}
\includegraphics[width=\textwidth]{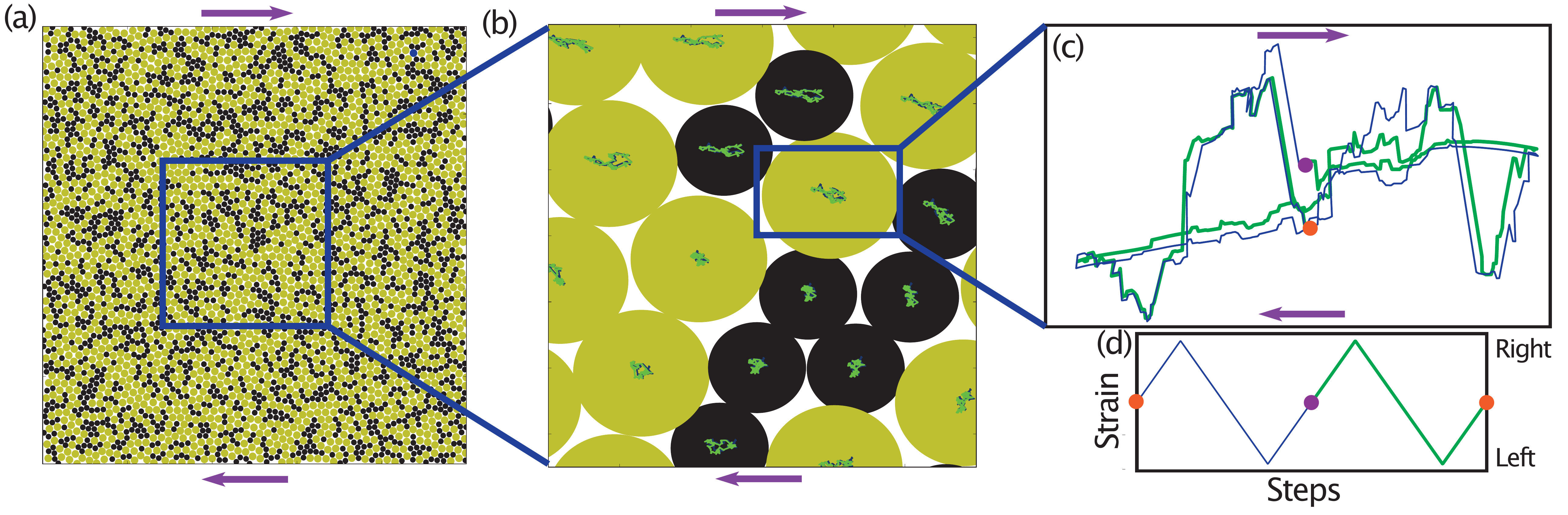}
\caption{
  Particle trajectories in an amorphous solid following a multi-periodic limit-cycle. (a) A system of 4096 particles subject to periodic shear. (b) A blow-up showing individual particles and the trajectories performed by their centers marked in blue and green, where blue represents the first cycle and green represents the second cycle. (c) A blow-up showing the trajectory of a single particle. During the first cycle the particle performs the blue trajectory, followed by the green trajectory during the second cycle.
(d) The strain as a function of simulation steps (quasistatic equivalent of time) in the cycle.
Reprinted under CC license from I. Regev {\it et al.}, Nature Commun. {\bf 6}, 8805 (2015).
}
\label{fig:M1}
\end{figure*}

\begin{figure}
\includegraphics[width=3.5in]{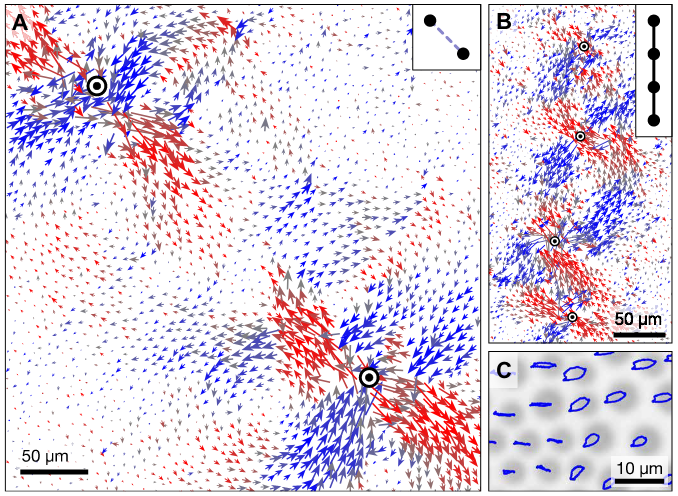}
\caption{
Experimental observation of interacting plastically deforming regions
or ``soft spots''
in the reversible state of
a jammed solid.
(a) Arrows indicate the magnitude
of the particle displacements and colors indicate the two principal shear
axes. (b) Cooperative interactions among multiple soft spots. (c) Trajectories
(blue) of individual particles in a reversible state where the particles
return to their initial positions after each cycle.
Reprinted under CC license from N.~C.~Keim and J.~D.~Paulsen, Sci.~Adv.~{\bf 7},
eabg7685 (2021).
}
\label{fig:10}
\end{figure}

As shown in Fig.~\ref{fig:M1} and Ref.~\cite{Regev13}, the reversible state 
does not have 
to recur during each drive cycle, but may instead recur after
multiple driving cycles. For example,
the pattern might repeat after two cycles, and 
in fact limit cycles of up to seven or more cycles
have been observed \cite{Regev13,Lavrentovich17,Keim21}. 
Multiple other studies found that periodically sheared
amorphous systems can organize to reversible states
with a varied number of limit cycles.
This behavior is reminiscent of the routes into chaotic states that arise
in low dimensional systems \cite{Lorenz63,Ott93,Tel08}.
Keim {\it et al.} \cite{Keim21}
showed that the multi-periodicity can be explained as resulting from interactions between soft-spots
such as the ones shown in Fig.~\ref{fig:10},
while Szulc et al. \cite{Szulc22} explained how oscillations in the activation thresholds of the soft-spots cause multi-periodicity.

\begin{figure*}
\includegraphics[width=\textwidth]{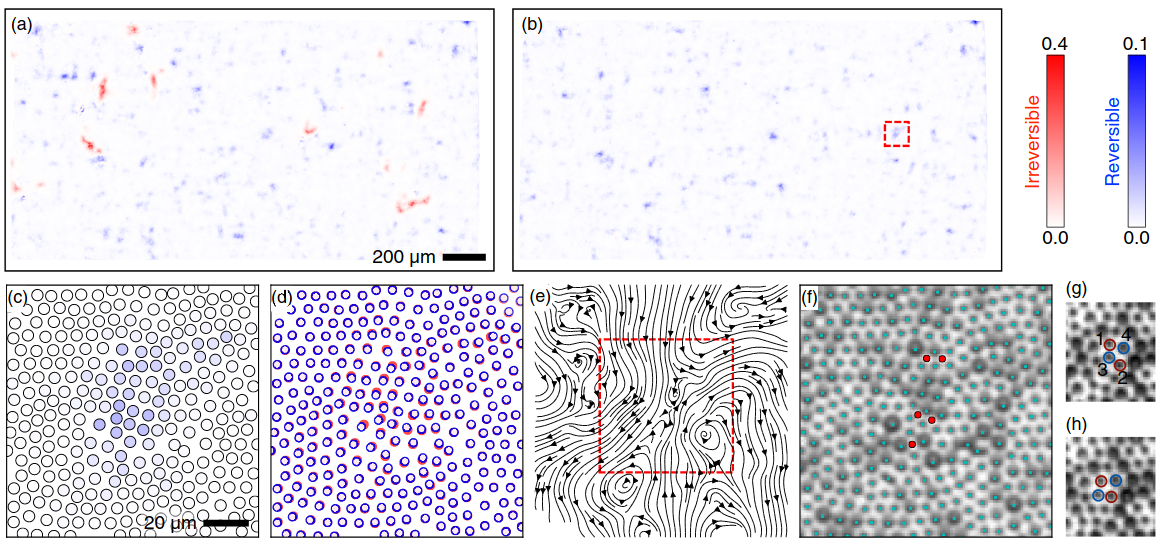}
\caption{Experimental measurements of local plastic deformation in a 2D jammed system. (a,b) Reversible (blue) and irreversible (red) motion after (a) 8 shear cycles and (b) 20 shear cycles showing that the system settles into a reversible state. (c) Detailed view of the residual displacement for the reversible cluster outlined with a red box in panel (b). (d) Relative maximum displacements of the particles in this cluster at the point of minimum shear (red) and at maximum shear (open blue circles). (e) The corresponding flow streamlines. (f, g, h) Images showing the details of a plastic rearrangement event in the same region.
Reprinted with permission from N.~C.~Keim and P.~E.~Arratia, Phys. Rev. Lett.
{\bf 112}, 028302 (2014).
Copyright by the American Physical Society.
}
\label{fig:9}
\end{figure*}

There have also been a number of experiments on
transitions into reversible plastic states in amorphous jammed systems. 
Keim {\it et al.} found that jammed systems can organize into
reversible plastic states \cite{Keim14}.
An example of this appears
in Fig.~\ref{fig:9}, which
highlights a detail of how the system organizes to
a state with reversible plastic events.
Nagamanasa {\it et al.} \cite{Nagamanasa14} also found 
a power law diverging time scale for
the transition to an irreversible state in a binary colloidal glass.  

In several works \cite{Priezjev16,Leishangthem17,Fan17,SchinasiLemberg20,Yeh20},
the effect of sample preparation on the potential energy in the steady-state
was considered. For sub-critical amplitudes, the steady-state potential energy of samples with different initial mean energies depends on the
initial conditions, but for post-critical initial conditions,
the steady-state potential energy is independent of the preparation protocol.
These studies also showed that for post-critical amplitudes, the potential energy increases up to saturation.
These observations were explained using models of dynamics on random energy landscapes
\cite{SchinasiLemberg20,Sastry21,Mungan21}.
Studies on ``ultra-stable'' glasses, or amorphous samples prepared using special quenching protocols, have shown that in these samples the transition from reversible to irreversible dynamics are abrupt and occur at critical amplitudes that depend on the preparation protocol \cite{Yeh20,Bhaumik21}. This is contrary to samples prepared using ``standard'' quenching protocols where the critical amplitude does not depend on sample preparation. Other studies have shown that within the reversible state,
although the system remains solid it becomes a softer solid, particularly when the reversible trajectories form loops \cite{Otsuki22}.

Efforts to model the R-IR transition in amorphous solids subject to cyclic shear have so far focused on using integer automata models that represent an amorphous solid as a lattice of soft-spots interacting by elastic interactions. 
Using such a model, Khirallah {\it et al.} \cite{Khirallah21} found that such systems indeed exhibit a transition
at a critical amplitude
from asymptotically periodic, where the system repeats after $n$ forcing cycles, to asymptotically diffusive dynamics.
They also found that, similar to what is observed in particle simulations, the transition between the reversible and irreversible phases is marked by a power law divergence in the number of cycles required to reach a reversible state.
The exponent observed in this case was
$\nu \approx 2.7$, consistent with the exponents found by Regev {\it et al.} \cite{Regev13}
and lending further support to the idea that the universality class
of R-IR transitions in amorphous solids differs from that of the dilute systems.
Khirallah {\it et al.} also observed that similar to the observations from particle simulations\cite{Fiocco13,Regev18}, 
the diffusion coefficient on the irreversible side of the transition increases as a power law in the magnitude of the strain amplitude, and showed that
there are still a large number of reversible plastic events that occur
within the irreversible state.
This work suggests that cellular automata models, which
are computationally faster than particle simulations and are easier to model theoretically,
can capture many of the relevant behaviors
at R-IR transitions, and that similar reversible plastic to
irreversible diffusive phases
could arise in other types of cellular automata models.

There are several possible avenues for continued research in the study of amorphous solids under cyclic shear. First, 
there are a variety of different amorphous systems in which R-IR transitions were not yet studied such as polymer glasses,
metallic glasses, gels, and emulsions. It is not clear if such systems will show the same transition and, if so, whether the transition would be of the same character. Second, the theoretical understanding of the transition in amorphous solids remains undeveloped. Although it is clear that in dilute systems, reversible dynamics arises due to the emergence of states where the particles are spaced in such a way that they are not interacting, it is not clear why in some amorphous configurations the interactions between soft-spots lead to irreversible dynamics whereas in others the same interactions allow for periodic states.
Third, it would be interesting to study
how the 
R-IR transition varies for different kinds of particles.
For example,
particles in
densely packed emulsions
can undergo a variety of shape changes under compression
or have very different types of elastic properties compared
to hard particles \cite{Mason06}.
It may be that the ability of
individual particles to distort
would introduce another form of dissipation that could
increase the range of reversible behavior;
however, such distortions could
also inject
additional degrees of freedom, giving more possible
ways for the packing to change and promoting irreversible motion.

\begin{figure}
\includegraphics[width=3.5in]{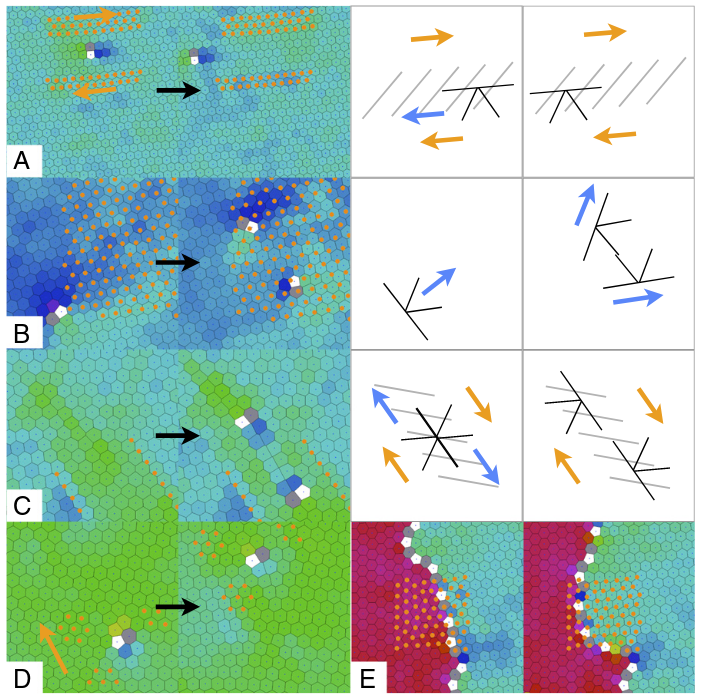}
\caption{Motion of dislocations and grain boundaries being controlled with optically induced ``topological'' tweezers in experiments on a colloidal assembly.
(a) Inducing glide with localized shearing. (b) Inducing climb by dilatation. (c) Dislocation fissioning through applied shear. (d) Glide of a dislocation that has been trapped by opposing shear stresses. (e) Moving a grain boundary by applying a potential that is commensurate with the lattice on one side of the boundary. From W.T.M. Irvine {\it et al.}, Proc. Natl. Acad. Sci. (USA) {\bf 110}, 15544 (2013).
}
\label{fig:11}
\end{figure}
        
Beyond disordered solids,
R-IR transitions
have also been studied 
in polycrystals \cite{Jana17,Ni19} and
point dislocation models \cite{Otsuki22}.
There have been only a few studies of  R-IR
transitions in crystalline systems \cite{Ni19},
but there are a variety of effects that could be studied in
ordered states,
such as the motion of grain boundaries
or of isolated defects such
as dislocation lines and disclinations.
In crystalline systems,
application of periodic driving could generate defects, leading
over time to work hardening and eventual failure;
however, there could be regimes in which the system reaches a
steady reversible state under cyclic driving.
This could be tested for 
crystalline systems found in soft matter, materials science,
and certain hard condensed matter systems such as
superconducting vortices or Wigner crystals.
In Fig.~\ref{fig:11}, dislocations in a 2D colloidal assembly are
manipulated using local stresses, shears, and dilatations \cite{Irvine13}.
One could consider applying local or global periodic driving
to such a system in order to determine whether
the motion of
the individual defects illustrated in 
Fig.~\ref{fig:11}(a,b,c,d) or of the grain boundary illustrated in
Fig.~\ref{fig:11}(e) is reversible.

\subsection{Magnetic Systems}

Hysteresis is frequently observed in condensed matter and
materials science, and the best known example is in
magnetism where cycling an applied field generates a hysteretic
magnetic response in the material
\cite{Chikazumi97}.
Hysteresis in magnetic systems is a result of disorder and exchange interaction between the spins of different atoms.
In the case of ferromagnetic interactions,
disorder can take the form of a local random field,
as in the random field Ising model (RFIM)
which has the Hamiltonian \cite{sethna1993hysteresis}:
\begin{equation}
H = -J\sum_{\langle i,j\rangle}s_is_j  - \sum_i h_is_i - hs_i
\end{equation}
where $s_i = \{-1,1\}$ is the direction of the $i$th spin, $J>0$ is a constant ferromagnetic coupling constant, $h_i$ is the random field and $h$ is an externally applied field.
Alternatively, there can be randomness in 
the effective spin-spin interactions,
as captured by the Edwards-Anderson (EA) spin-glass model \cite{edwards1975theory}:
\begin{equation}
H = -\sum_{\langle i,j\rangle}J_{ij}s_is_j - hs_i
\end{equation}
In this case the coupling constant $J_{ij}$ is a random variable that can be both positive and negative, leading to geometric frustration.
Models with ferromagnetic interactions where $J>0$
always reach a limit cycle after
a transient of two cycles or less due to the ``no-passing''
property first proved by Middleton \cite{middleton1992asymptotic,sethna1993hysteresis}. For this reason, such systems cannot have a R-IR transition. 
Models with couplings that can be both positive or negative, such as a spin-glass, can have long transients and thus, in principle, can have
both reversible and irreversible behavior. In the case of the EA spin-glass and related systems, each spin has only two states, which 
may hinder
the emergence of completely irreversible dynamics.
At the thermodynamic limit,
however, the transients may become infinitely long,
and the system can then become effectively irreversible.

\begin{figure}
\includegraphics[width=3.5in]{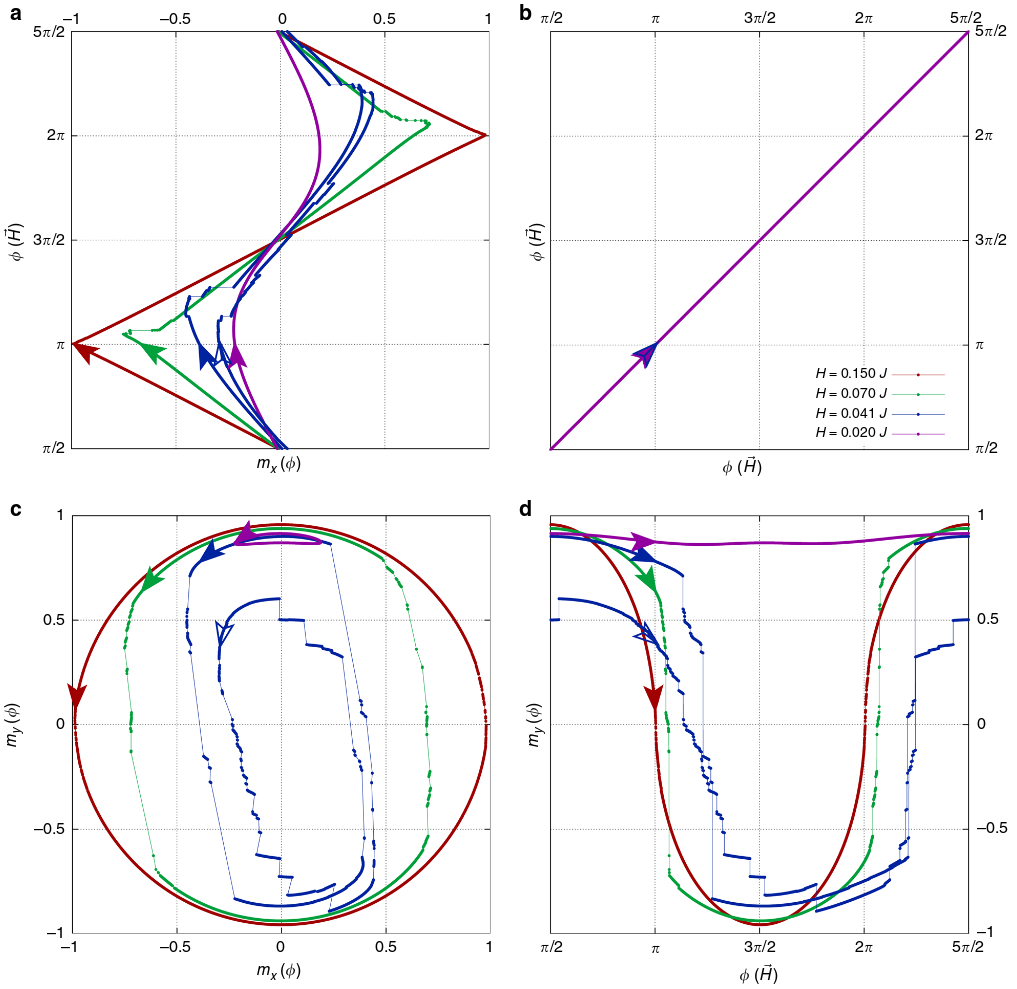}
\caption{
Cyclic driving of uniaxial random field XY models with disorder.
(a) The driving field angle $\phi$ versus
the magnetic response in the $x$ direction $m_x$ for increasing
driving strengths, where purple is the lowest drive and red is the
highest drive.
(b) A plot of
$\phi$ versus $\phi$ indicating that the applied field is rotated
counterclockwise. (c) The $y$ direction magnetic response $m_y$
plotted versus $m_x$. (d) $m_y$ versus $\phi$.
Reprinted under CC license from S. Basak {\it et al.}, Nature Commun. {\bf 11}, 4665 (2020).
}
\label{fig:21}
\end{figure}

Basak {\it et al.} \cite{Basak20}
considered cyclic driving of uniaxial random field XY models with disorder.
They found that for increasing field amplitude,
the system goes from an Ising ferromagnetic state
to a paramagnetic state that does not repeat,
as shown in Fig.~\ref{fig:21}.
In the reversible case, the spin patterns 
repeat after $n$ cycles,
and the number of cycles to reach a 
reversible state increases as the critical point is approached.
The plot of the $y$ and $x$ magnetizations $m_y$ versus $m_x$ in
Fig.~\ref{fig:21}(c) 
indicates that there is an initial transient
before the system settles into a period-2 limit cycle.
This implies that the
magnetic system can organize into
a repeatable pattern spanning one or multiple periods, similar
to what was found in the
amorphous solids discussed previously.

Uniaxial random field XY models can be applied to many systems,
including Josephson junctions \cite{LacourGayet74},
superfluids in a uniaxially stressed aerogel \cite{Aoyama06},
uniaxially stressed 2D Wigner crystals \cite{Nelson79},
the half-integer quantum Hall effect,
and the graphene quantum Hall ferromagnet \cite{Abanin07}.
Electron nematics \cite{Carlson06,Carlson11} are also promising.
Each of these are
candidate systems in which to look for R-IR transitions
under periodic driving.

\section{Future Directions in Condensed Systems}

\subsection{Commensurate-Incommensurate Systems}

\begin{figure}
\includegraphics[width=3.5in]{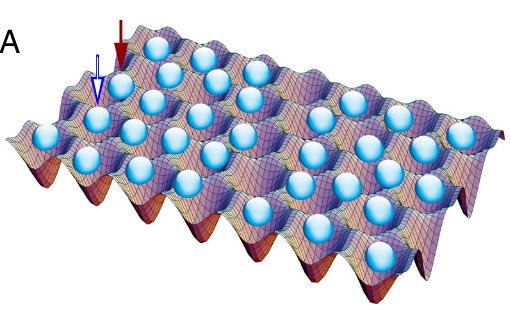}
\caption{An illustration of a 2D version of a Frenkel-Kontorova model
consisting of colloidal particles interacting with a periodic
substrate.
From A. Vanossi {\it et al.}, Proc. Natl. Acad. Sci. (USA) {\bf 109}, 16429 (2012).}
\label{fig:12}
\end{figure}

Another class of systems that are good candidates for examining
R-IR transitions
is that in which
commensurate-incommensurate transitions
can occur \cite{Bak82,Coppersmith82,Braun98,Reichhardt17}.
These systems can be described
in terms of interacting particles on a periodic substrate,
where the ratio of the number of particles $N$ to the number of
substrate minima $N_s$ is given by the
filling ratio $f = N/N_{s}$.
For example,
Fig.~\ref{fig:12}(a) shows a schematic of
charged colloidal particles
interacting with a 2D periodic egg-carton substrate
at a filling close to $f=1$ \cite{Vanossi12}.
Commensurate conditions can arise for fractional matching
when $f=n/m$ with integer $n$ and $m$
\cite{Reichhardt17}. The most
celebrated example of this
is the Frenkel-Kontorova model
for one-dimensional (1D) chains of elastically coupled particles on a
substrate \cite{Braun98};
however, commensurate-incommensurate transitions
arise across a remarkable variety of hard and soft
condensed matter systems in both one and two dimensions.
Specific systems include
atomic ordering on surfaces \cite{Bak82,Coppersmith82},
cold atoms in optical traps \cite{Bloch05},
vortices in nanostructured superconductors \cite{Harada96,Reichhardt98a},
vortices in Bose-Einstein condensates \cite{Tung06},
and colloidal systems \cite{Vanossi12,Brunner02,Bohlein12,McDermott13a}.
In quantum systems, there can be transitions from
commensurate Mott phases to incommensurate
superfluids \cite{Fisher89,Bakr10}.
For most of the systems listed above,
the particles are not strictly elastically 
coupled but can undergo exchanges with one another,
permitting plastic deformations, defect generation,
and phase slips to occur.

\begin{figure}
\includegraphics[width=3.5in]{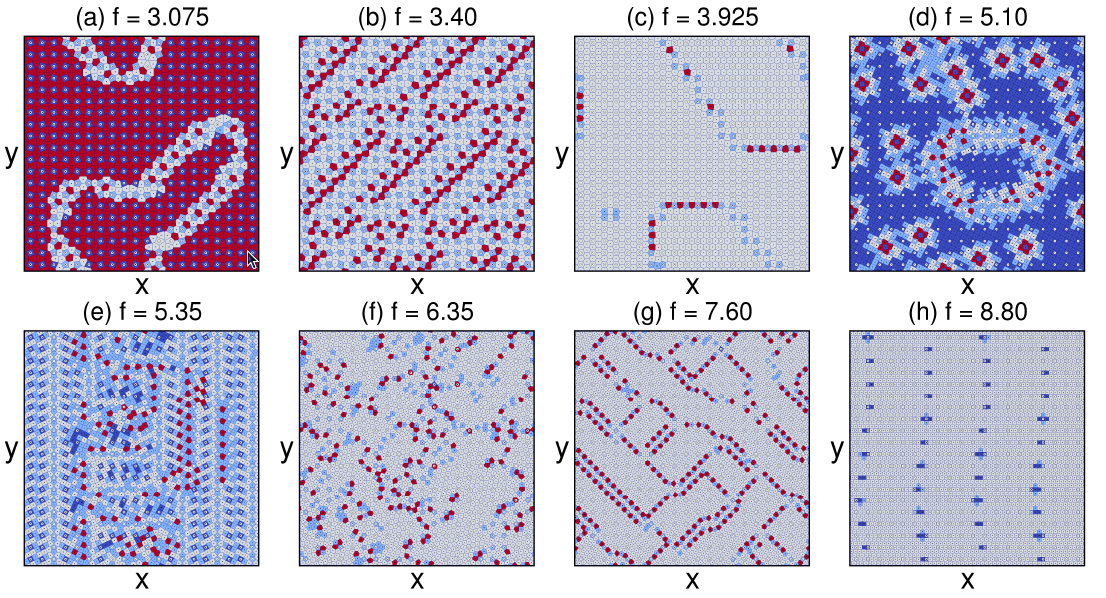}
\caption{Voronoi tessellations of colloidal particles interacting with a
periodic substrate for different nearly commensurate fillings ranging
from $f=3.075$ to $f=8.80$. Dark blue polygons are fourfold coordinated,
light blue polygons are fivefold coordinated, gray polygons are sixfold
coordinated, and red polygons are sevenfold coordinated. A variety of
grain boundary and dislocation structures appear.
Reproduced from Ref.~\cite{McDermott13} with permission from the
Royal Society of Chemistry.
}
\label{fig:13}
\end{figure}

In typical commensurate-incommensurate systems,
the particles are largely
localized under commensurate conditions,
while for slightly incommensurate states there are kinks or anti-kinks
that are more mobile
than the particles and
that depin first under an external drive \cite{Reichhardt17}.
At more strongly incommensurate states,
the system becomes increasingly disordered and can
exhibit glass-like behavior.
Since the amount of order can be tuned by changing the filling,
it would be possible to examine R-IR transitions
for varied fillings.
For example, under commensurate conditions
the system may be strongly reversible for a range of drives,
whereas incommensurate fillings
might
either organize to a reversible state similar to
what is found in the sheared systems
or remain in an irreversible
state.
There could be a critical drive amplitude marking the R-IR transition,
or there could be fillings for which the system becomes fluid-like
and is always irreversible. 
Whether the response of a commensurate-incommensurate
system
is reversible or irreversible may also depend on the
type of drive applied. 
In addition to their relevance to a wide class of systems,
another advantage of studying commensurate-incommensurate systems 
is that a variety of distinct types of defect structures can
be realized, ranging from individual solitons or kinks and domain walls to
strongly amorphous phases. 
To illustrate why commensurate-incommensurate systems could
show different reversible or irreversible behaviors,
Fig.~\ref{fig:13} 
shows that a wide range of defect morphologies emerge for
colloids on periodic substrates at different incommensurate fillings
\cite{McDermott13}.
It would be interesting to understand whether the grain boundaries
and isolated defects that emerge in a system such as this one would
exhibit reversible or irreversible dynamics under periodic driving,
and whether a possible R-IR transition
would resemble that
found in dilute
or amorphous systems,
or whether it would fall into an entirely new universality class that
has not yet been observed for R-IR transitions.
Such a study would relate to
the broader question of
the nature of R-IR
transitions in interacting soliton systems.

\subsection{Local Driving}

\begin{figure}
\includegraphics[width=3.5in]{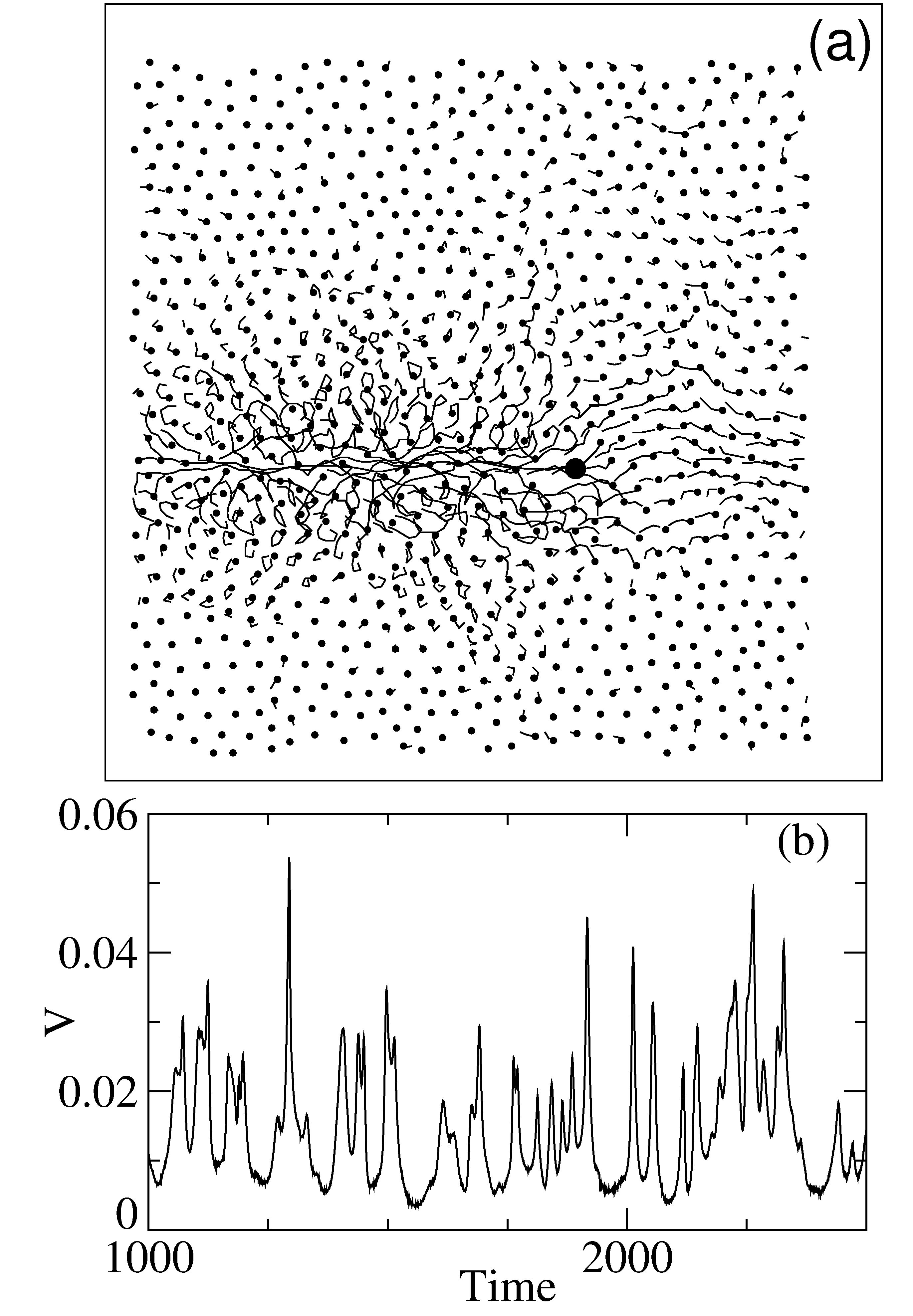}
\caption{(a) Particle locations (small black dots) and trajectories (lines) in a simulation of a single probe particle (large black dot) being dragged through a glassy background of bidisperse colloidal particles.
(b) Time series of the velocity of the probe particle in panel (a) shows jumps corresponding to plastic rearrangement events.
Reprinted with permission from M.~B.~Hastings {\it et al.}, Phys. Rev. Lett.
{\bf 90}, 098302 (2003).
Copyright by the American Physical Society.
}
\label{fig:14} 
\end{figure}

In the systems discussed up to this point,
the driving is applied globally
via shearing, a current, or a field;
however, it is also possible to subject a system to
local driving. In soft matter, such driving is referred to as
active rheology, where a single particle is
dragged through a
background medium in order to create measurable perturbations
\cite{Hastings03,Habdas04,Squires05,Candelier10,Yu20}.
Studies of this type have been used
to examine changes in the drag \cite{Squires05,Yu20} and
fluctuations \cite{Hastings03,Candelier10,Illien14}, as well as to
determine whether there is
a threshold force needed
for motion of the probe particle \cite{Hastings03,Habdas04,Senbil19}
as the system
passes through different types of glassy or jamming transitions.
Figure~\ref{fig:14}(a) shows an example of a single probe particle driven
through a glassy background of bidisperse colloidal particles, and the
corresponding velocity time series
contains a
series of jumps that indicate the occurrence of
plastic events
in the medium surrounding the probe particle \cite{Hastings03}.
Similar active rheology techniques have also been used
to study plastic rearrangements in crystalline systems
\cite{Reichhardt04a,Dullens11}.

\begin{figure}
\includegraphics[width=3.5in]{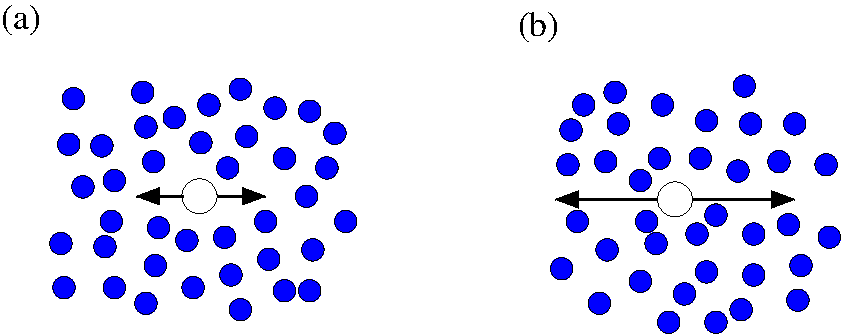}
\caption{Schematic of how local periodic driving could be used to detect
an R-IR transition with a local probe by oscillating the probe at low (left)
and high (right) amplitudes.
}
\label{fig:15}
\end{figure}

In hard condensed matter systems,
local driving can be achieved using
a magnetic force tip or scanning tunneling microscope for
superconducting vortices \cite{Straver08,Auslaender09} or an optical probe for
both superconducting vortices \cite{Veshchunov16}
and magnetic skyrmions \cite{Hanneken16}.
Other possible methods include
injecting a current at a single
localized spot or
applying time dependent inhomogeneous fields.
Most studies of active rheology have used dc driving,
but ac driving could also be applied.
Local driving has been applied to
plastic to elastic transitions
as well as situations where the probe particle particle creates local
plastic deformations in the system. Similar techniques
could be used to explore
local R-IR. For example, if
the probe particle is periodically driven at different driving
amplitudes, as illustrated schematically in Fig.~\ref{fig:15},
the system could organize to an
elastic reversible state with no plastic deformation,
to a state with reversible plastic events or loop reversibility,
or to a state that is continuously fluid-like or irreversible.  
It would be possible to measure whether
a reversible or irreversible interface extends out from the region
containing the driven particle, or if there is a fluctuating or
continuous border surrounding the local probe region.
To test for memory effects,
the ac drive could be applied in one direction until the system
reaches a reversible state, and could then be rotated into a new
direction to see whether the system still remains reversible.
Such an approach would also be useful for
understanding how the dissipation or drag is
affected when the system transitions from
an irreversible to a reversible state.
In work examining sedimenting colloids
under a periodic
shear \cite{Wang22}, the presence of a density gradient made it possible
to determine that
the denser regions are irreversible
while the less dense regions are reversible,
and that the reversible and irreversible regions are separated
by a well defined coexistence front.

\subsection{Complex Interactions}

\begin{figure}
\includegraphics[width=3.5in]{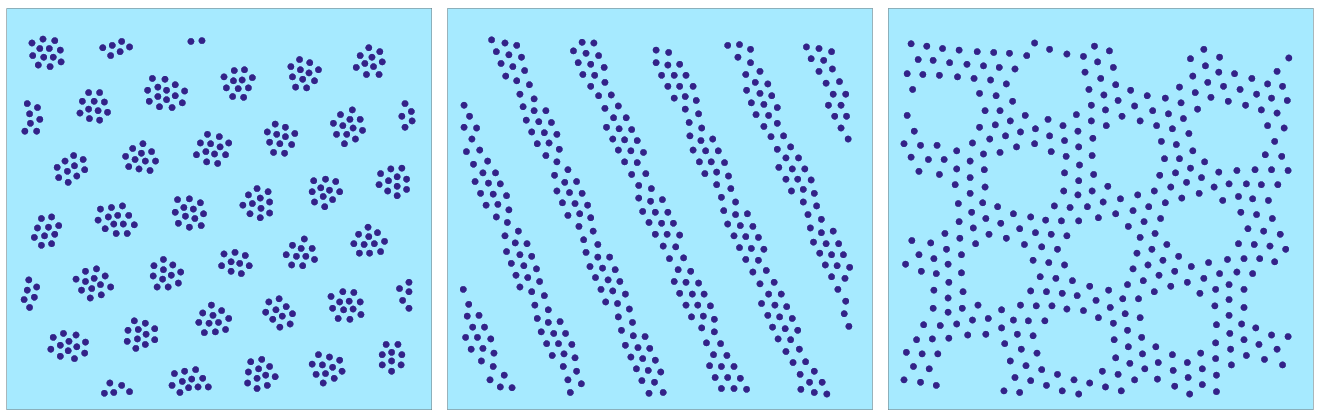}
\caption{Example of clump (left), stripe (center), and
void (right) patterns that form for particles with competing
long range repulsion and long range attraction from a system such as
that described in Ref.~\cite{Reichhardt10}.
}
\label{fig:16}
\end{figure}

Up until this point, R-IR transitions have been studied
only for systems with relatively simple short range 
interactions.
There are, however, numerous systems in which
more complex interactions
appear, such as long range repulsion and short range attraction, frustrated
interactions, and interactions with multiple length scales.
When these types of complex interactions are present,
pattern formation or phase separation
often occurs, as found in many soft matter
systems \cite{Seul95,Malescio03,Reichhardt03a,Reichhardt10}
as well as in
hard condensed matter systems such
as electron liquid crystals \cite{Fradkin99,Cooper99,Gorez07},
doped Mott insulators \cite{Kivelson98},
charge ordering in Jahn-Teller systems \cite{Mertelj05}
and magnetic domains \cite{Seul92}. 
In Fig.~\ref{fig:16} we show some examples of pattern
formation in a system
with multiple length scale interactions, which cause
both short range crystalline order and larger scale ordering
to occur.
It would be
interesting to test whether systems of this type
could exhibit R-IR transitions
under periodic driving.
For soft matter systems, this could be achieved
through shearing,
while for hard
condensed matter systems
such as electron liquid crystals, the driving could take the form
of an oscillating current \cite{Cooper99,Gorez07}. 
Other systems with long
range interactions where some form of periodic driving can be imposed
include
dusty plasmas or charged dust particles that form microscale
crystals \cite{Thomas94,Chiang96}.
Such systems could be periodically sheared or 
driven with other methods \cite{Gogia17}. 

Due to the multiple length scales that are present in the interaction,
it is possible that individual particles might undergo irreversible
deformations even as the larger scale pattern remains reversible,
producing a situation with
microscopic irreversibility but macroscopic reversibility.
If this were the case, there might be multiple R-IR transitions as well as
the possibility for
complex memory formation.

\subsection{Quasi-Reversibility and Universality Classes}

The systems described so far are in either a reversible or an irreversible
state during one or multiple drive cycles. Another possibility is that
even in the absence of long time diffusion,
the system could follow one of several
available loop paths, where on any given cycle the particular path that is
chosen is selected randomly.
This could occur if
the system shows only a local ergodicity
or if there are regions where the particles
can locally explore many different possible states
but local fluctuation paths remain confined,
so that the particles are unable to diffuse over long times.
There could also be
glassy reversibility.
In many of the systems described so far, the irreversible 
states were characterized by observing diffusive behavior of the
particles; 
however, there could be regimes of very long
time diffusion in which the system is trapped in one cyclic orbit for
long times before making a rare jump to a new orbit.
Glassy behaviors of this type
could arise when thermal fluctuations become important.
For example, the system could be
in a reversible regime for $T = 0$, but become irreversible
at finite $T$ when it can occasionally
thermally hop to different periodic orbits.
Finally, there could be orbits that are quasi-periodic in time,
in analogy to orbits that are quasi-periodic in space. 

Another possibility is
smectic-like irreversibility in which
there is long time diffusion in one direction
but not in the perpendicular direction.
In driven systems with quenched disorder, moving
smectic states have been observed in which the
diffusion is finite in one direction but not in
the other \cite{Reichhardt17}.
For colloids or 
amorphous solids under a periodic drive,
1D reversibility of this type
could occur for anisotropic particles or
layered systems.
There could then be two irreversible transitions that occur separately
for each direction.
It would be interesting to understand whether
there are only a few types of distinct universality classes for
R-IR transitions,
or whether multiple universality classes could be
observed by changing the nature of the particle-particle interactions
or the driving.

\section{Vortices, Skyrmions, and Other Hard Condensed Matter Systems}

There are many other
systems containing assemblies of particles
that can be periodically driven.
These include cases in which
the particles
are coupled to random or periodic substrates \cite{Reichhardt17,Fisher98},
as found 
for sliding friction \cite{Vanossi13}, vortices in type-II superconductors
\cite{Bhattacharya93,Koshelev94},
sliding charge density waves \cite{Balents95,Danneau02},
colloids \cite{Pertsinidis08},
pattern forming systems \cite{Reichhardt03a},
active matter \cite{Morin17}, election liquid crystals \cite{Cooper03},
and Wigner solids \cite{Williams91}.
The driving can be applied uniformly
using an applied current, an electric field, 
or a magnetic field.
If there is no quenched disorder present and 
the drive is uniform, the particles simply translate back and forth;
however, when disorder is present,
certain particles can become trapped
while other particles move past them, leading to plastic events.
Under an increasing dc drive,
these systems typically exhibit a pinned phase,
a plastically deforming phase,
and a moving crystal or moving smectic state.
If the
disorder is weak,
the system depins elastically
without the creation of
topological defects.

\begin{figure}
\includegraphics[width=3.5in]{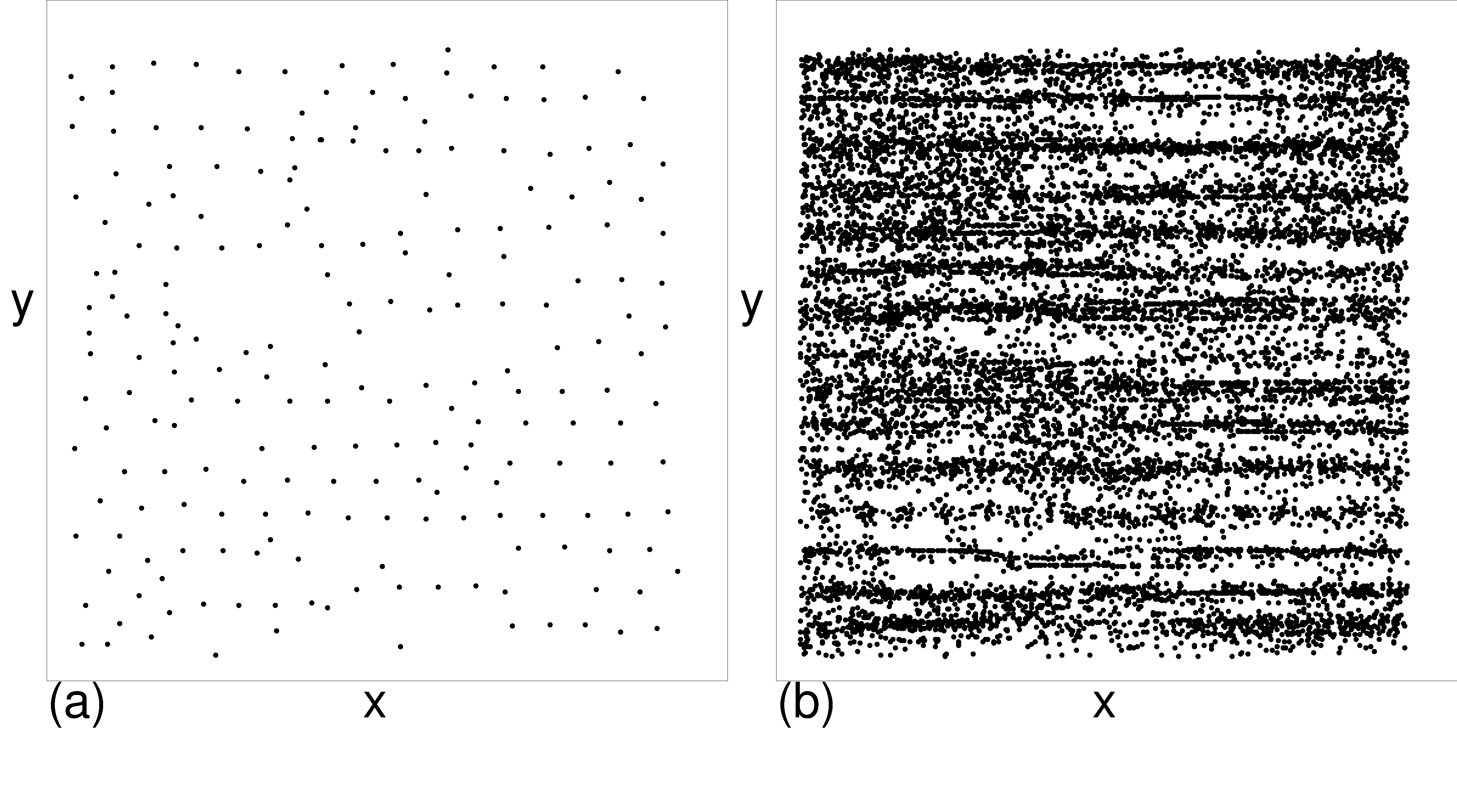}
\caption{Stroboscopic snapshots of the positions
during 25 cycles of superconducting vortices
cyclically driven over a random landscape.
(a) The reversible regime for small ac drive amplitudes.  
(a) The irreversible regime for large ac drive amplitudes.  
Reprinted with permission from N.~Mangan {\it et al.}, Phys. Rev. Lett.
{\bf 100}, 187002 (2008).
Copyright by the American Physical Society.
}
\label{fig:17}
\end{figure}
  
Mangan {\it et al.} \cite{Mangan08} performed numerical simulations
of superconducting
vortices in a 2D system with random disorder and ac driving where the
driving period, vortex density, and pinning strength are varied.
Figure~\ref{fig:17}(a) shows the vortex positions at the end
of each cycle after 25 cycles under a drive for which each vortex moves a
distance
of $45\lambda$ during a single cycle, where $\lambda$ is the London
penetration depth. Here the system is in a reversible regime.
In Fig.~\ref{fig:17}(b), when the driving amplitude is increased so that
each vortex moves a distance of
$160\lambda$ during every cycle,
the vortices do not return to the same positions
and the system is irreversible.
For the reversible state,
the diffusion is zero
and the time series of
the average vortex velocity
repeats the same pattern during every drive cycle, which would produce
a periodic voltage signal if measured experimentally.
In the irreversible regime,
the vortices exhibit a finite diffusion and the
voltage signal is chaotic.
Okuma {\it et al.} \cite{Okuma11}
studied R-IR
transitions experimentally by shearing
superconducting vortices in a Corbino geometry, and
mapped
out the phase diagram for
the transition from reversible to irreversible flow
as a function of vortex displacement per cycle.

\begin{figure}
\includegraphics[width=3.5in]{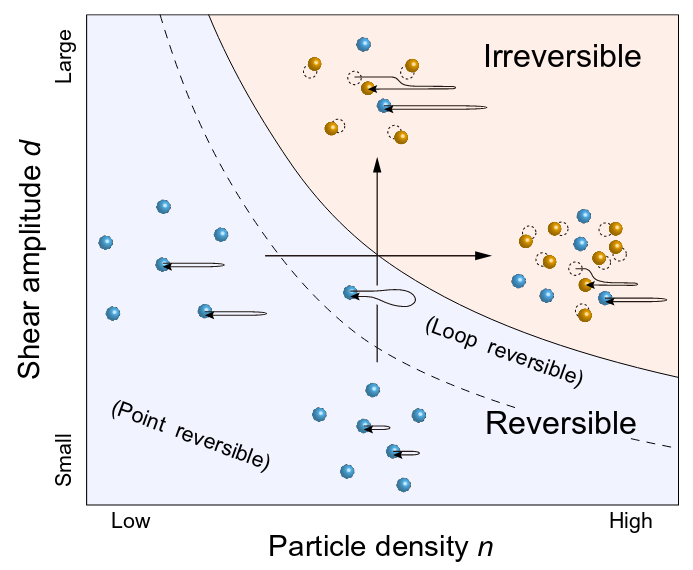}
\caption{
Schematic illustration of R-IR phases in a superconducting vortex
system as a function of shear amplitude $d$ versus vortex or particle
density $n$. Point reversible states appear for low densities or small
shears, while a region of loop reversibility emerges prior to the
transition to irreversible behavior at high densities and large shears.
Reprinted under CC license from S. Maegochi {\it et al.}, Sci. Rep. {\bf 11}, 19280 (2021).
}
\label{fig:18}
\end{figure}

Additional work has since been performed 
\cite{Maegochi21} to more clearly show
that superconducting vortices can also exhibit point reversible states
similar to the random organization
found for dilute colloids,
loop reversible states similar to
those observed in amorphous systems,
and irreversible states, as illustrated
in Fig.~\ref{fig:18}.
At low densities where the vortices are far apart, the
behavior resembles that of
a dilute colloidal system, and
point reversible states emerge,
while at higher density where interactions between the vortices
become important,
the system shows loop
reversibility.
For large densities and large
driving amplitudes, the system is irreversible.
Ref.~\cite{Maegochi21} also showed that there is
a diverging time scale for the system to settle into a
reversible or irreversible state.
The power law divergence
reported in Fig.~4 of Ref.~\cite{Maegochi21}
is similar to that
found in the random organization system
and has an exponent $\nu = 1.33$
that is
close to the value expected
for directed percolation \cite{Takeuchi09}.
Even though the vortices
can exhibit collective effects due to their
longer range interactions,
the exponents observed
are consistent with the dilute colloidal system rather
than with amorphous solids,
where exponents closer to $\nu=2.6$ are found.
This could be due to the
nature of the plastic events that occur in the different systems.
There has also been work
showing evidence for R-IR transitions in
ac driven superconducting vortices in linear
geometries \cite{Pasquini21}.

Magnetic skyrmions are another particle-like magnetic texture that
have many similarities to
superconducting vortices in that they can be driven by
an applied current \cite{Schulz12,Reichhardt15}.
An important distinctive feature is that skyrmions have a strong
gyrotropic or Magnus force.
Simulations of skyrmions under ac
drives with quenched disorder \cite{Brown19} showed an R-IR
associated with a diverging time scale where the exponent is
close to $\nu=1.29$.
In this system, the Magnus term enhances the irreversible behavior by
increasing the number of dynamically accessible orbits. This is in
contrast to the behavior of an overdamped system, such as strongly damped
skyrmions and vortices in type-II superconductors. It has been shown that
in the overdamped limit, there appear to be two different R-IR
transitions, 
since
the diffusion constant first drops to zero in the direction perpendicular
to the drive, followed by a regime in which the behavior is reversible
both parallel and perpendicular to the drive.
This suggests that the number of possible R-IR
transitions may
depend on the effective dimensionality of the system.
Experimental results that seem consistent with the predictions
of Ref.~\cite{Brown19} were found
in the vortex system \cite{Maegochi19}.

\section{Future directions for systems with quenched disorder}

\begin{figure}
\includegraphics[width=3.5in]{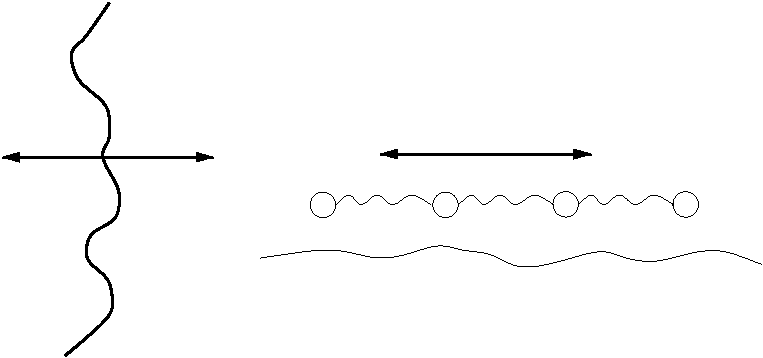}
\caption{
Schematic of other elastic systems that could exhibit R-IR
transitions
under ac driving (arrows), such as a moving contact line (left) or a sliding
charge density wave (right).
}
\label{fig:19}
\end{figure}

R-IR transitions under periodic driving could be
explored in
charge density wave systems \cite{Balents95,Danneau02},
which often exhibit narrow band noise or temporal ordering
that may be the hallmarks of  reversible behavior.
It would be interesting to understand if there is also an irreversible
regime and if there is
a diverging time scale for reaching the reversible regime.
This is of particular relevance since recent work on classical time 
crystals has suggested
that sliding charge density waves under ac driving are examples of
discrete
time crystals \cite{Yao20},
which we discuss further in a later section.
Other elastic systems that can be driven periodically
include magnetic domain walls \cite{Han09},
moving contact lines \cite{Blake79},
depinning of interfaces \cite{Leschhorn93a},
and slider block models \cite{Fisher98}.
Examples of possible materials science systems
are ac driven grain boundaries \cite{Turnbull51}, 
twin planes \cite{Wang17},
and dislocation patterns \cite{Bako08,Zhou15}, as illustrated
in
the schematic of Fig.~\ref{fig:19}.
It is possible that systems of this type would always organize
to a reversible state;
however, there could be different kinds
of reversible states separated by transitions with diverging
time scales.
For example,
there could be a critical point
separating point reversible states and loop reversible
states.

\subsection{Systems With Avalanches and Noise} 

\begin{figure}
\includegraphics[width=3.5in]{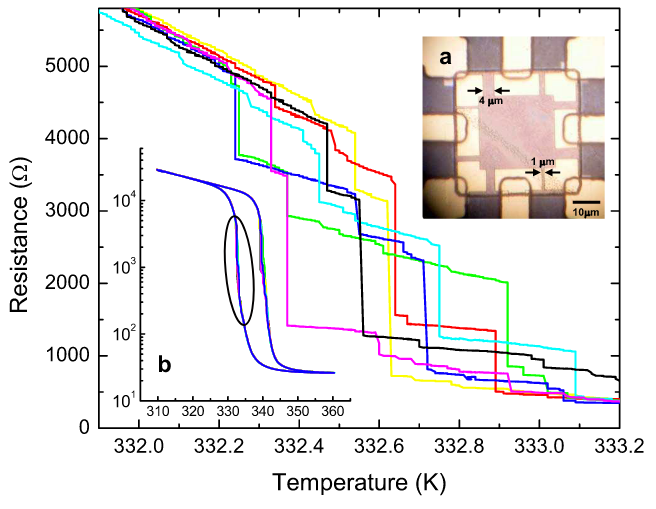}
\caption{
Experimentally measured hysteresis loops in a VO$_2$ sample
at the metal-insulator transition. (a) Illustration of the sample
geometry showing that eight devices are present on a single
sample; the widths of two of the devices are marked with arrows.
(b) Full hysteresis loop as a function of resistance versus
temperature. Main panel: Consecutive resistance versus temperature
cycles shown zoomed in on the portion of the full hysteresis loop
circled in panel (b). Jumps occur that are not repeatable from cycle
to cycle.
Reprinted with permission from A.~Sharoni {\it et al.}, Phys. Rev. Lett.
{\bf 101}, 026404 (2008).
Copyright by the American Physical Society.
}
\label{fig:22}
\end{figure}

Further directions include studying systems with hysteresis that
exhibit repeating avalanches 
to seek diverging scales under repeating
hysteretic cycles.
Figure~\ref{fig:22}
illustrates experimentally measured
avalanches across the metal-insulator transition
of VO$_2$ \cite{Sharoni08}.
In systems of this type, it would be possible to perform
repeated cycles to determine
whether the same resistance values recur, and whether there is a finite
number of cycles that must be performed before the system
reaches a repeating pattern. 

In many condensed matter systems, noise is generated under the application
of a current or field.
This noise can take the form
of avalanches or crackling noise \cite{Sethna01,Weissman88},
telegraph noise \cite{Levy91a},
switching \cite{Hayakawa21},
narrow band \cite{Bhattacharya87,Reichhardt97}, or broad band noise
\cite{Reichhardt97,Marley95},
and can be characterized by using the
power spectrum or second spectrum \cite{Weissman88}.
Under applied periodic driving, it would be interesting to
explore whether the
system exhibits the exact same noise pattern,
and if so,
whether a finite number of cycles must be performed before
the system reaches a reversible state. 

\subsection{Astrophysical Systems}

\begin{figure}
\includegraphics[width=3.5in]{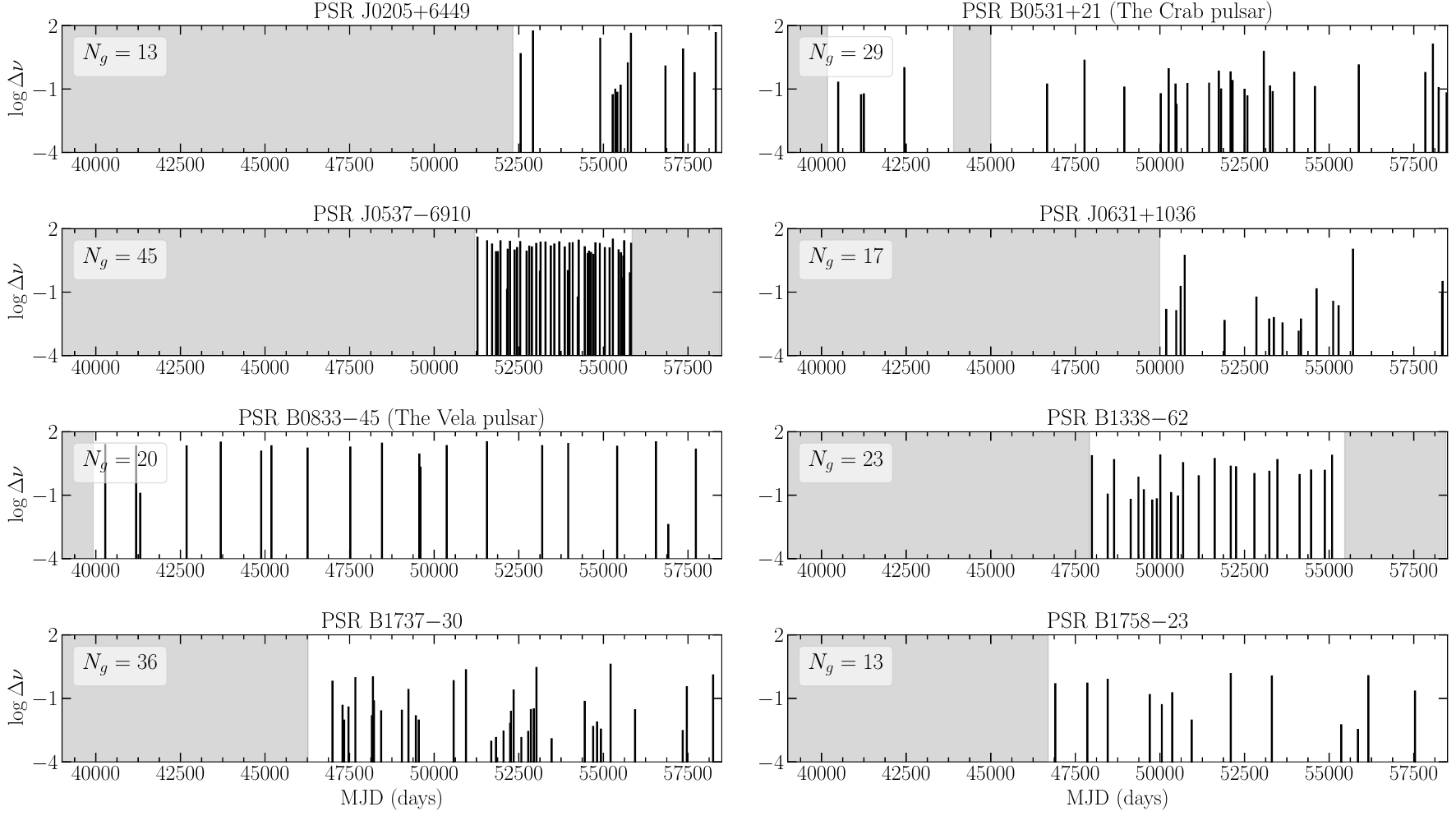}
\caption{Observations of glitches $\delta \nu$ (in $\mu$Hz)
of the rotation frequencies $\nu$ for different pulsars
as a function of time in units of
the modified Julian date (MJD).
The total number of glitches observed for each pulsar is listed as $N_g$,
and windows of time in which no observations were made for at least 3 months
are shaded in gray.
Reprinted under CC license from J.R. Fuentes {\it et al.}, Astron. Astrophys. {\bf 630}, A115 (2019).
}
\label{fig:23} 
\end{figure}

Another set of systems that
can be considered to have periodic driving and
that also
exhibits avalanches or bursts
are pulsar glitches in neutron stars
\cite{Melatos08,Howitt18,Espinoza11,Carlin18} and brightness variability for
certain stars \cite{Sheikh16}.
Here the periodic driving
arises from the rotation of the stars.
When a pulsar glitch occurs, there is a shift $\delta \nu$ in the rotation
frequency $\nu$ of the pulsar.
In general, glitching pulsars fall
into two classes: those
with Poisson-like waiting times \cite{Melatos08,Howitt18,Espinoza11},
and those with unimodal or quasiperiodic  waiting times
\cite{Espinoza11,Carlin18}. 
Figure~\ref{fig:23} shows some example time series
of pulsar glitches exhibiting
different levels of periodicity. 
It is possible that the glitches could be a sign that over time the star
is
organizing to a more periodic state,
and if so, a reversible state could emerge in which there is
a power law distribution of periodic waiting times.
It would be interesting to understand whether there
could be a transition from chaotic
glitch intervals to periodic or quasiperiodic glitches, and whether
a diverging time scale appears in these systems.
Sheikh {\it et al.} \cite{Sheikh16}
measured avalanches in the brightness variability of stars
and found scaling exponents that suggest
the system may be near a nonequilibrium critical point.
A system of this type might be
irreversible on one side of the nonequilibrium point
and reversible on the other side.
	
\begin{figure}
\includegraphics[width=3.5in]{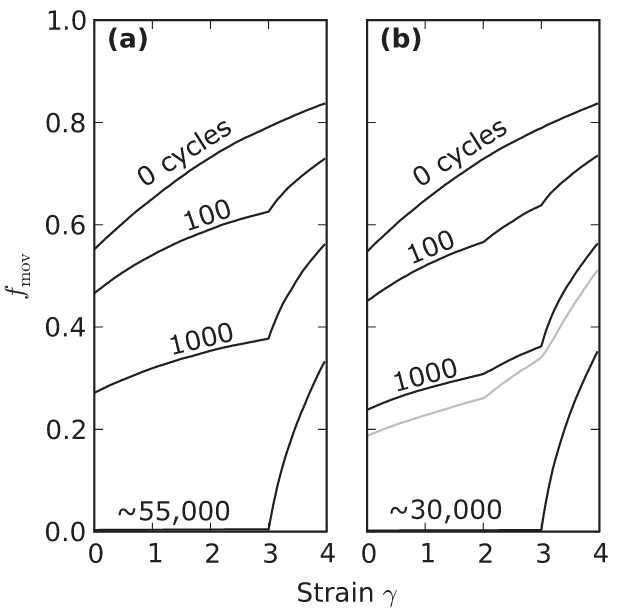}
\caption{The fraction of moving or irreversible particles $f_{\rm mov}$ versus
the trial strain level $\gamma$ for different numbers of cycles in
a numerical model of sheared colloidal particles, demonstrating
the emergence of memory.
(a) Application of a single shear level $\gamma_1$ gives a kink in the
trial strain $\gamma$ at $\gamma=\gamma_1$; reversible behavior eventually
emerges for $\gamma<\gamma_1$ but the system remains irreversible for
$\gamma>\gamma_1$.
(b) Application of a smaller shear level $\gamma_2$ for five cycles
followed by application of a larger shear level $\gamma_3$ for one cycle
gives the trial strain level $\gamma$ a memory of the value of both
$\gamma_2$ and $\gamma_3$. When the system organizes to a reversible state,
only memory of the larger strain level $\gamma_3$ is retained.
Reprinted with permission from N.~C.~Keim and S.~R.~Nagel, Phys. Rev. Lett.
{\bf 107}, 010603 (2011).
Copyright by the American Physical Society.
}
\label{fig:24}
\end{figure}

\section{Memory effects} 
Sheared colloidal systems
can organize to a reversible state
when the applied strain
$\gamma_i$ satisfies $\gamma_i < \gamma_{c}$,
where $\gamma_c$ is the critical stain below which the system
is always reversible.
If the strain amplitude is
changed so that $\gamma_{i} > \gamma_{c}$, reversible behavior
cannot appear.
Paulsen {\it et al.} \cite{Paulsen14} considered
a colloidal random organization model in which a
memory effect emerges
for strains smaller
than $\gamma_{c}$.
In Figure~\ref{fig:24}(a),
a numerical study
\cite{Keim11} of a system with a critical strain level
of $\gamma_{c} = 4.0$.
showed that when a training shear pulse amplitude of
$\gamma_{1} = 3.0$ is applied, a ``reading'' of the system with a trial
strain level of $\gamma$ produces a signature
or memory of $\gamma_1$ in the form of a cusp.
It is also possible to store the values of two distinct
strain levels, 
as shown in Fig.~\ref{fig:24}(b)
for $\gamma_2 = 2$ and $\gamma_2 = 3.0$, where
there are now two kinks that appear as the trial strain level
is varied.
This implies that multiple memories can be
storied in these systems as long
as the strain amplitude remains
below the critical stain where
the system becomes irreversible and loses all memory.
Fiocco {\it et al.} \cite{Fiocco14} studied a protocol similar to
that used by Keim {\it et al.} \cite{Keim11} but employed
an amorphous solid in which
the particles always remain in contact. They find that
the same memory effect shown in Fig.~\ref{fig:24}(a) can be achieved by
using a training shear amplitude.

\begin{figure}
\includegraphics[width=3.5in]{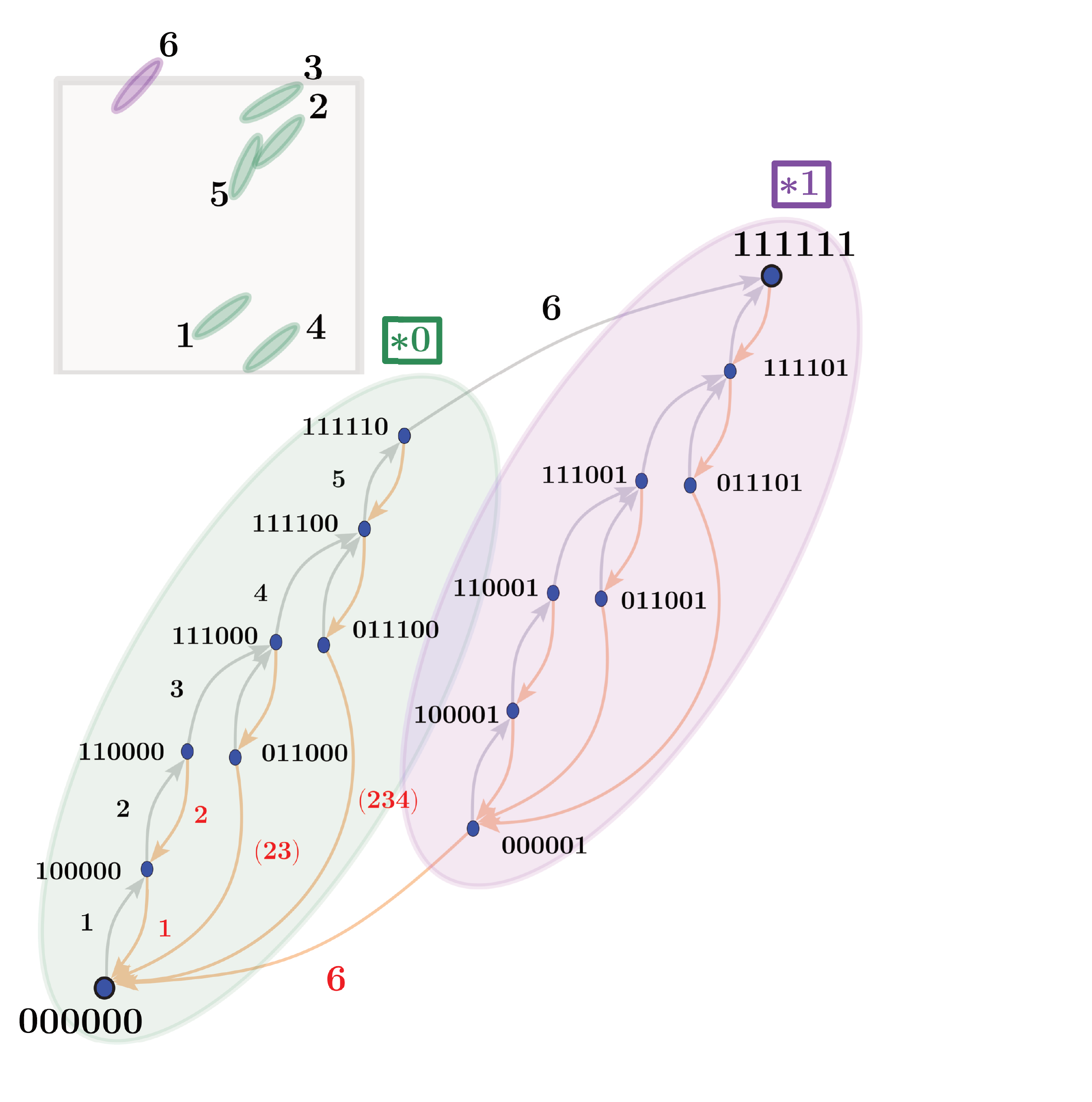}
\caption{A network representation of the transitions between different states
of a model of an amorphous solid subjected to cyclic shear deformations.
The inset shows the soft-spot locations and numbers.
Each state is represented by a binary number where 1 or 0 in the second digit represents a state where soft spot number 2 is ``switched on'' or ``switched off,''
respectively.
Arrows represent transitions due to an increase (black) or decrease (red)
in the strain causing soft-spot switching. The green and purple shaded regions are limit-cycles with soft-spot number 6 switched off or on, respectively.
Reprinted with permission from M. Mungan {\it et al.}, Phys. Rev. Lett.
{\bf 123}, 178002 (2019).
Copyright by the American Physical Society.
}
\label{fig:25}
\end{figure}

Mungan {\it et al.} \cite{Mungan19} showed that the configurations of an amorphous solid sheared along a fixed plane can be represented by a directed graph where nodes represent stable particle configurations and arrows represent transitions between them due to plastic events.
Figure~\ref{fig:25} shows an example of such a network where all the transitions are reversible. In each configuration a soft spot
is switched on or off.
Regev et al. \cite{Regev21} further showed that the strongly connected components of the graphs, representing clusters of configurations where every two configurations (nodes) are reachable by a path of plastic deformation, include all the possible limit-cycles of the system.
Since transitions between different strongly
connected components are irreversible, as the strain magnitude increases
to a level
close to the R-IR transition,
irreversible transitions become more frequent and the strongly connected components become smaller and can contain only small limit-cycles, such as the
yellow arrows in Fig.~\ref{fig:25}.
Some open questions
include whether introducing a change in the direction of shearing
or driving in this regime could reduce the persistence of the memory.
Although memory has been studied in sheared colloids and sheared
amorphous solids,
less is known about whether
similar memory effects can occur in other systems.
For example, Dobroka {\it et al.} \cite{Dobroka17} found
that kinks similar to those found
by Paulsen {\it et al.} \cite{Paulsen14}
can occur for periodically driven superconducting vortices.

\begin{figure}
  \includegraphics[width=3.5in]{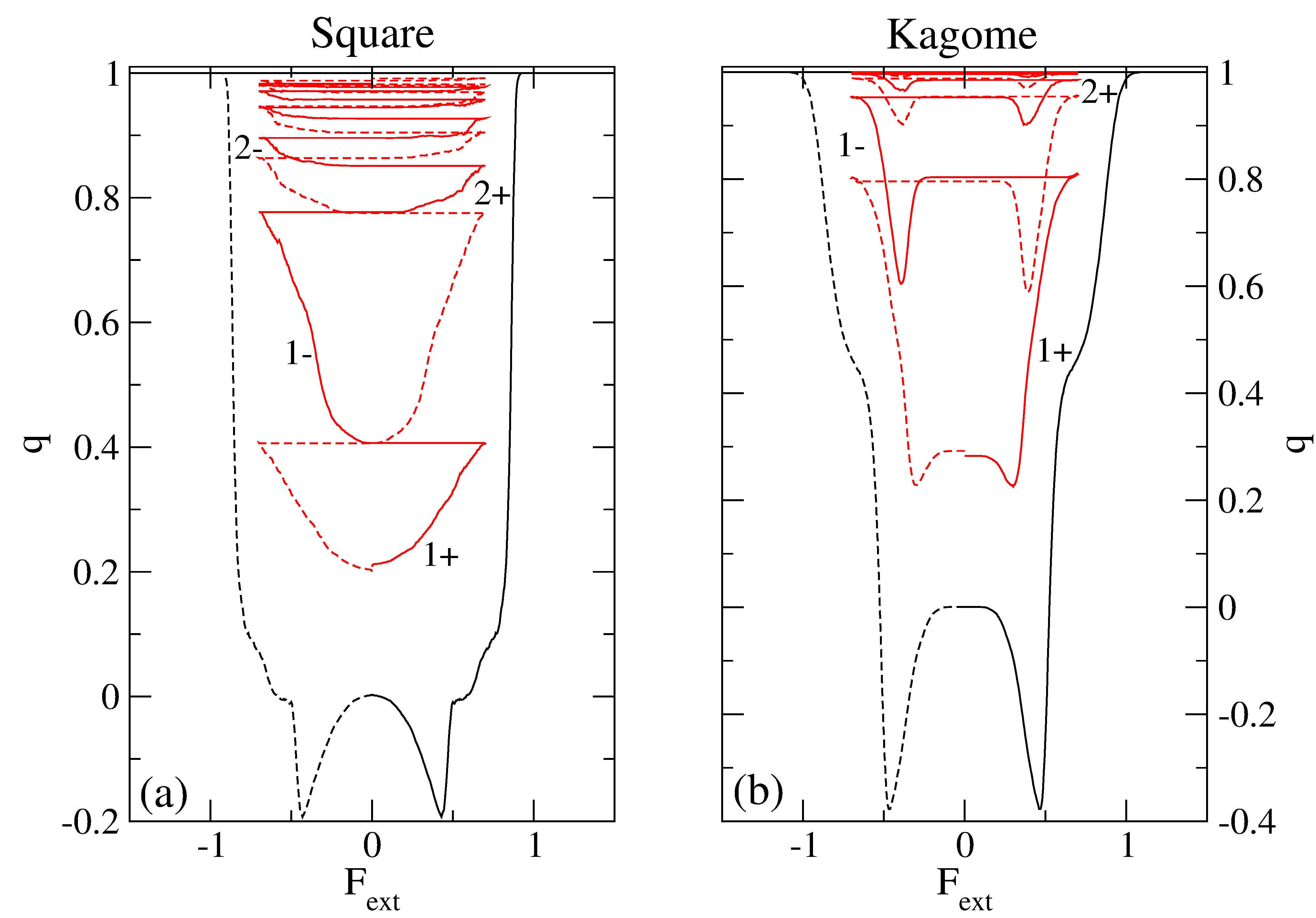}
\caption{
Return point memory for minor loops in an artificial spin ice system constructed
from colloidal particles.
When $q=1$, the system reaches a reversible state. For both the
(a) square ice and (b) kagome ice geometries, $q$ increases as the number
of minor loop cycles performed increases (numbers), indicating the
emergence over time of reversible behavior.
Reprinted with permission from A. Lib{\' a}l {\it et al.}, Phys. Rev. E
{\bf 86}, 021406 (2012).
Copyright by the American Physical Society.
}
\label{fig:20}
\end{figure}

Memory can also be associated with the hysteresis that appears in
magnets,
semiconductors \cite{Shaw92},
and metal-insulator systems \cite{Liang20} under oscillating voltage,
and at dynamic transitions in magnetic systems subject to fast oscillations where the hysteresis is with respect to dynamic variables \cite{Chakrabarti99}.
For example, in the phenomenon of
return point memory \cite{Pierce05,Katzgraber06}, 
spin configurations 
are examined at the end of each minor loop cycle
to see if the configurations fully overlap from cycle to cycle according
to an overall function.   
In periodically driven colloidal spin ice systems \cite{Libal12}, the
spin overlap function $q$ was examined for repeated minor loops.
A value $q = 1.0$ means that
exactly the same effective spin configuration appears during each
cycle, indicating reversible behavior. 
In Ref.~\cite{Libal12},
the system did not adopt the same
spin configurations for every cycle,
but as shown in Fig.~\ref{fig:20}, $q$ increases with time until saturating
at a reversible
state after a fixed number of cycles,
similar to the random organization
observed for dilute colloids. In addition,
including quenched disorder increased the memory of the system.

\section{Future Directions for Time Crystals and Quantum Systems} 

Another
promising system in which to seek R-IR transitions
is time crystals,
which have been proposed for both quantum \cite{Wilczek12}
and classical systems \cite{Shapere12}.
In a time crystal, the lowest energy state of the system is periodic
not only in space but also in time.
Shortly after the initial proposals,
it was recognized that true time crystals 
as originally envisioned cannot occur under conditions of
strict equilibrium
\cite{Bruno13,Watanabe15};
however,
the time crystal concept
has now generated a wealth of ideas for
creating time periodic systems
that could arise under
nonequilibrium conditions \cite{Sacha18,Else20}, such as
periodic driving.
A key feature of
a time crystal is that the system can exhibit
subharmonics of the oscillatory driving
\cite{Else20,Kehmani16,Yao17} (multi-periodic response).
Yao {\it et al.} \cite{Yao20} have noted
that there are a number of classical systems that also
show subharmonic entrainment,
including Faraday waves \cite{Cross93} and
predator-prey models \cite{Vasseur09},
phase locking in driven charge density wave
systems \cite{Gruner88} or Josephson junctions \cite{Lee91a},
and superconducting vortex
\cite{Reichhardt00b} or magnetic skyrmion \cite{Reichhardt15b}
motion on periodic
pinning arrays,
meaning that these systems
could be examples of Classical Discrete Time Crystals (CDTCs).
In quantum systems, time crystals have been studied
by measuring subharmonic entrainment
in unitary many-body systems or Floquet systems \cite{Yao17}.
In some cases, time crystals
can arise in quantum systems
where many-body localization can prevent thermalization \cite{Sacha18,Else20}.
Up until now,
work on time crystals has focused on
identifying systems that support time crystals;
however, the
dynamics of how a system can organize into a time
crystal state or the general phase diagram
near the boundary from chaotic to time crystal behavior
has many similarities to the R-IR transitions discussed above.

\begin{figure}
  \includegraphics[width=3.5in]{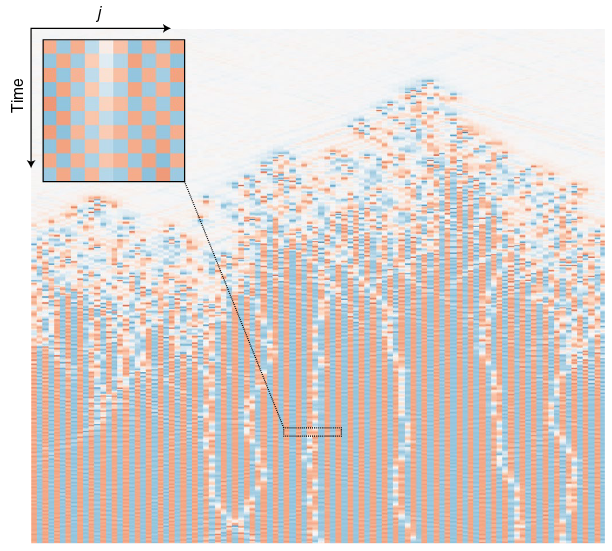}
\caption{A stroboscopic image as a function of time versus position $j$ of a
  one-dimensional system of coupled oscillators.
Oscillators that have a position coordinate $q_i$ less than zero are
red, those with $q_i>0$ are blue, and those with $q_i \approx 0$ are white.
Period-doubled oscillations emerge from the uniform state over time,
forming a
Classical Discrete Time Crystal (CDTC). 
The behavior is similar to the organization of
a fluctuating system into a reversible
state as shown earlier.
Reprinted by permission from: Springer Nature,
N.~Y.~Yao {\it et al.}, ``Classical discrete time crystals,''
Nature Phys. {\bf 16}, 438 (2020).
  }
\label{fig:26}
\end{figure}

An example of a classical time crystal that appears close to systems
in which R-IR transitions can happen appears
in the work of Yao {\it et al.} \cite{Yao20},
where a one-dimensional array of coupled oscillators was  studied at
finite temperature.
Figure~\ref{fig:26} shows a
stroboscopic view of the position coordinate $q_j$ of each oscillator
under periodic driving
as a function of time.
In the initial state, the system is 
disordered or fluctuating,
but over time it organizes into an antiferromagnetic
repeating state.
This is reminiscent of an initially irreversible state organizing
over time into a reversible state.
The reversibility does not have to appear after a single driving cycle;
instead, the system can repeat after two or more driving cycles.
It would be interesting to see whether there
is a time scale for the organization into the
time periodic or CDTC state, and if so, whether this time diverges
as a function of the driving parameters, similar to what is found
for the R-IR transition
in colloidal systems. 
The work of Yao {\it et al.} \cite{Yao20} can also be
applied to the much broader class of systems in higher dimensions and
at zero temperature,
such as systems that exhibit phase locking effects
where nonlinearities give rise to higher order harmonic phase locking.
There could be a variety of other systems on periodic flashing substrates
that could organize to
reversible states over time,
such as choreographic colloidal time crystals that have
a liquid-like state
and a time ordered state \cite{Libal20}.

As discussed in a previous section, the reversible states can exhibit memory
and training effects, so similar phenomena
along with memory encoding studies could be explored in time crystal
systems, where
much more complicated periodic driving protocols can be employed.
For example, the system might
organize to a CDTC for one set of
driving frequencies, but if
additional frequencies were added,
the system could organize into a new time crystal state.
The question would be whether, if the system is first trained with the
initial set of frequencies, these
frequencies could be retained in the time crystal
state formed with additional driving frequencies.
Although we have focused on classical time crystals,
similar R-IR transitions could
arise in periodically driven quantum time crystals,
such as periodically driven trapped
atomic ions \cite{Zhang17},
driven spins in diamond \cite{Choi17},
and arrays of superconducting qubits \cite{Mi22}.

\section{Complex Networks and Dynamical Systems} 

In general, there are many other coupled systems
that can organize into synchronized states.
Systems that exhibit ergodicity breaking would also be candidates
for study in the framework of R-IR transitions.
For example, dynamical many-body coupled oscillators
are known to show transitions to synchronized states \cite{Strogatz15}.
Other systems include particles coupled
to or flowing on a complex network \cite{Cameron14} 
under periodic driving.
In this case, the particles could form repeating loop paths
on the network in the reversible state, such as those illustrated
in Fig.~\ref{fig:10}(c).
Such systems could include those for which
particles or carriers of information
cannot pass though each other
and the network is rigid.
Other systems
such as those with
nonreciprocal
interactions \cite{Fruchart21} or odd viscosity  \cite{Soni19}
show transitions from chaotic to
ordered edge states,
and could organize to patterns
containing ordered loop currents or flocking.

Complex networks and dynamical systems
exhibiting synchronization
arise across many biological \cite{OKeefe17,Feillet14}, 
robotic \cite{Wang21},
social \cite{Shahal20},
and economic systems \cite{Selover99}.
It is known that these
systems can all enter chaotic or strongly fluctuating states;
however, there could be situations in which, when some type of
periodic driving is applied,
time repeating states could emerge similar to the reversible states
we have discussed  above.
Here, similar types of memory formation could arise,
or there could be some form of power law divergence
in a time scale near a critical point.

\section{Summary}
We have given an overview
of the recent work examining
transitions from chaotic irreversible states to  time periodic or reversible
states for systems under periodic driving.
In these systems, the locations of the particles are compared
to their positions on the previous
driving cycle.
For irreversible states, the particles do not return
to the same positions, and over multiple cycles 
they exhibit diffusive
motion away from their initial positions.
In a reversible state, the
particles return to the same positions after one or an integer number
of driving cycles, and there is no long time diffusion.
Reversible-irreversible (R-IR) transitions were initially studied
for periodically sheared dilute colloids,
in which a process termed random organization produces
reversible states
where particle-particle collisions no longer occur,
while the irreversible states have continuous collisions. 
R-IR
transitions have also been studied
in strongly interacting systems such as amorphous solids
and jammed systems.
Another hallmark of the R-IR transition
is that there is a critical drive or 
density above which  the
system remains in an irreversible state,
and there is a power law divergence on either side of the transition for the
time required for the system to settle into a
steady irreversible state or
organize into reversible motion.
In dilute systems,
the R-IR transition is consistent with directed percolation.
For amorphous solids, similar R-IR behavior is found
but the transition appears to fall into a different universality class,
and for dense systems, the critical amplitude
coincides with the yielding transition. 
On the reversible side of the R-IR transition
it is possible to store a series of memories
by applying training pulses.
Such memory effects have been observed in both dilute
and dense systems.
R-IR transitions
have also been studied in solid state systems such as periodically 
driven superconducting vortices and
magnetic skyrmions, both of which show
similar behavior to that found in dilute sheared colloidal systems.

We highlight how the general features of the
irreversible to reversible dynamics transition could be applied 
to much broader classes of soft matter, hard matter and dynamical systems.
For example, many solid state systems show hysteretic behavior,
and systems of this type would be fertile ground for studying
transitions to reversible states under repeated cycling.
The impact of
repeatable noise, avalanches,
or return point memory could be studied,
as well as the number of cycles required
to reach the reversible state, which
could show critical behavior similar to that found in colloidal systems. 
Such studies could be performed for
magnetic systems, metal-insulator transitions, charge ordering systems,
and semiconductors. Other classes of system
in which to look for R-IR transitions include
commensurate-incommensurate systems, frustrated systems,
crumpled and corrugated sheets \cite{Shohat22,Bense21},
and even active matter systems.  
These systems may also be able to store more complex memories.
For example, amorphous solids or colloids could
be sheared with a periodic but increasingly complex protocol
to see if the system can still reach a reversible state.
We also discuss some other possible states
that are not time periodic
but for which there is no long time diffusion, so that
quasiperiodic dynamics can be explored, as well as systems that
do not follow the same path on each cycle
but can still only trace out a limited number of cycles,
which could arise for frustrated states. 

We discuss how the R-IR framework could be applied
to classical and quantum discrete time crystals where the
transitions to time periodic or harmonically entrapped states could be
an example of an R-IR transition.
Power law divergences in time could appear
for the formation of time crystal
states.
Other systems to consider include more general cyclic systems
such as pulsars, coupled oscillators,
social systems, economics, biological systems, and  
particle flow dynamics on complex networks where the reversible
state could arise through the formation of loop currents.

\acknowledgments
This work was carried out under the auspices of the 
NNSA of the 
U.S. DoE
at 
LANL
under Contract No.
DE-AC52-06NA25396.
IR was supported by the Israel Science Foundation through Grant No.~1301/17.
KD thanks NSF for support through the {\it Expeditions in Computing} Program
NSF IIS-2123781.

\bibliography{mybib}
\end{document}